\documentclass[10pt,amsmath,amssymb,nofootinbib,twoside,twocolumn,superscriptaddress,floats,floatfix,aps,prd,preprintnumbers]{revtex4-2}
\usepackage[british]{babel}
\usepackage{xcolor}
\usepackage{booktabs}
\usepackage{latexsym}
\usepackage{amssymb}
\usepackage{amsmath}
\usepackage{mathrsfs} 
\usepackage{graphicx}
\usepackage{tabularx}
\usepackage{cancel}
\usepackage{hyperref}
\hypersetup{
    colorlinks=true,
    linkcolor=black,
    citecolor=blue,
    filecolor=black,
    urlcolor=blue,
    pdfauthor={A. Bonanno, M. Cadoni, M. Pitzalis, and A. P. Sanna},
    pdftitle={Effective Quantum Spacetimes from Functional Renormalization Group}
}
\usepackage[capitalise]{cleveref}
\usepackage[utf8]{inputenc}
\usepackage[normalem]{ulem}
\crefname{plural}{Eqs.}{Eqs.}
\Crefname{plural}{Eqs.}{Eqs.}

\crefformat{plural}{#2eqs.~(#1)#3}
\Crefformat{plural}{#2Eqs.~(#1)#3}

\newcommand{\dd}{\text{d}}

\begin{document}

\title{\bf Effective Quantum  Spacetimes  from  Functional Renormalization Group}

\newcommand{\UniCa}{\affiliation{Dipartimento di Fisica, Universit\`a di Cagliari, Cittadella Universitaria, 09042 Monserrato, Italy}}
\newcommand{\INFNCa}{\affiliation{INFN, Sezione di Cagliari, Cittadella Universitaria, 09042 Monserrato, Italy}}
\newcommand{\INAFCt}{\affiliation{INAF, Osservatorio Astrofisico di Catania, via S.Sofia 78, I-95123 Catania, Italy}}
\newcommand{\INFNCt}{\affiliation{INFN, Sezione di Catania, via S.Sofia 78, I-95123 Catania, Italy}}

\author{Alfio~Bonanno}
\email{alfio.bonanno@inaf.it}
\INAFCt\INFNCt

\author{Mariano~Cadoni}
\email{mariano.cadoni@ca.infn.it}
\UniCa\INFNCa

\author{Mirko~Pitzalis}
\email{mirko.pitzalis@ca.infn.it}
\UniCa\INFNCa

\author{Andrea~P.~Sanna}
\email{asanna@dsf.unica.it}
\UniCa\INFNCa

\begin{abstract}
Using the Functional Renormalization Group approach we  construct effective quantum spacetime geometries by self-consistently deforming the classical Schwarzschild-de Sitter black-hole solution. This involves studying  how quantum corrections, driven  by the running of the Newton's and cosmological constants modify the solution across the infrared and ultraviolet regimes. We show that these quantum modifications  replace the Schwarzschild singularity with a milder conical one. Moreover, two new features emerge in the ultraviolet regime. First, we identify a phase transition between Anti-de Sitter/de Sitter spacetime occurring when the object's mass exceeds a first critical threshold. Second, we predict the formation of horizons once the object's mass exceeds a second threshold. Both thresholds are of the order of the Planck mass. Finally, we investigate the role of the anomalous dimension in the conformal sector of the theory.
\end{abstract}

\maketitle
\tableofcontents

\section{Introduction}

In recent years, modern General Relativity has reached the status of a precision science. The forthcoming statistical data from gravitational-wave events produced by black-hole mergers will soon allow for precise predictions in the strong-field regime. However, despite significant progress in understanding the observable universe, the physics of black-hole interiors and the nature of cosmological singularities remain poorly understood. We still lack a definitive theory of black hole evaporation and a widely accepted theory of quantum gravity capable of addressing fundamental questions related to gravitational collapse.

Nevertheless, recent efforts to understand gravity within the framework of standard quantum field theory have yielded a clearer understanding of the vacuum's polarizability properties in the deep ultraviolet regime. In quantum electrodynamics (QED), vacuum polarization and the corresponding screening of the electric charge are important predictions of quantum field theory (QFT). The Coulomb potential becomes ``dressed'' by the Uehling term, which encodes the photon's self-energy within the photon propagator. This term can also be derived from the Euler-Heisenberg effective Lagrangian, which requires ultraviolet regularization, resulting in scale dependence. As it is well known, the renormalization group improvement for the running gauge coupling \( e^2 \rightarrow e^2(k) \), with \( k \propto 1/r \), encapsulates the key quantum fluctuation effects on the electric field produced by a point charge, governing the screening behavior at small distances \cite{Dittrich:1985yb}. Similarly, it is also well understood that both the qualitative and quantitative aspects of asymptotic freedom in quantum chromodynamics (QCD) can be explained as an effect of gluon self-coupling, which causes the vacuum to behave like a paramagnetic substance \cite{Nielsen:1980sx}.

It is not difficult to imagine the same mechanism operating in gravity \cite{Polyakov:1993tp}: as gravity is always attractive, a larger cloud of virtual particles strengthens the gravitational force. As a result, Newton's constant \( G \) should be \emph{antiscreened} at short distances, as in QCD. This effect implies that the dimensionless coupling constant \( g(\ell) = G(\ell) /\ell^2 \) approaches a finite, non-zero limit at short distances:
\begin{equation}
\label{as}
\lim_{\ell \rightarrow 0} g(\ell) \propto g^\ast \neq 0\, .
\end{equation}
Here, \( G \) scales as \( \ell^2 \) in accordance with its natural dimensions. This fact has far-reaching implications. 

A theory in which the dimensionless coupling constant approaches a non-Gaussian (non-vanishing) fixed point (NGFP) in the short-distance limit, as suggested in (\ref{as}), is termed \emph{asymptotically safe}, in contrast to the more familiar asymptotic freedom, where the coupling constants vanish at the fixed point. In fact, despite the similarities between non-abelian gauge theories and gravity, detailed investigations of the scaling behavior of Newton's constant became possible  only after the introduction of the effective average action (EAA) and its functional renormalization group equation (FRG) \cite{Reuter:1996cp}. Subsequent works based on the Wilsonian renormalization group (RG) approach have demonstrated that the theory's ultraviolet critical manifold is indeed governed by an NGFP with a finite number of relevant directions~\cite{Codello:2007bd,Falls:2014tra}.

The deep physical mechanism underlying the NGFP was finally clarified in \cite{Nink:2012vd}, where it was shown that above three spacetime dimensions, the gravitational antiscreening occurring in quantum gravity is due to a strong dominance of paramagnetic interactions over diamagnetic ones, which favor screening. As a consequence, spacetime can be interpreted as a polarizable medium with a "paramagnetic" response to external perturbations, very similar to the vacuum state of Yang-Mills theory.

This framework directly leads  to the notion of effective quantum geometries, where classical solutions are replaced by effective geometries featuring a running of Newton gravitational constant. This approach has produced the first static regular black hole within the asymptotic safety scenario~\cite{Bonanno:2000ep}. In this RG-improved quantum geometry, the Schwarzschild singularity at the origin  of the radial coordinate is removed and replaced with a de Sitter (dS) patch. 
The removal of the classical singularity due to quantum effects is a quite appealing feature from a phenomenological perspective. In fact, recent investigation have shown that, unlike the case of Schwarzschild black hole, the mass of non-singular black holes couple to the large scale cosmological dynamics~\cite{Cadoni:2023lum,Cadoni:2023lqe}.  Notably, the observational signature of this effect do not depend on whether the quantum effects that remove the singularity have  Planckian or super-Planckian origin.

Clearly, an approach based on the  notion of effective quantum geometry is capable of capturing only qualitative features of the RG flow near the NGFP, where all relevant scales can be neglected. However, there is no accepted solution for the structure of the EAA at $k=0$ which  should, in principle, reproduce the full effective action (EA). A pragmatic view holds that, in a real physical situation, the RG flow towards the infrared (IR) would be halted below a certain mass scale in the presence of matter. Conversely, in the deep ultraviolet (UV) regime, the scaling predicted by the RG equations should capture the essential physics.

Recent calculations, based on the so-called fluctuation approach~\cite{Pawlowski:2020qer} have  shown that, at least for pure gravity, the UV running of the physical Newton constant $G_N$ as a function of the external momenta $p^2$ at $k=0$, is indistinguishable from the running of $G_{k}$ near the NGFP (see, in particular, Fig. 4 of Ref.~\cite{Bonanno:2021squ}).

The interesting question is whether the approach of the effective quantum geometry  approach can be systematically extended to other relevant coupling constants.  Efforts to include  the running of the cosmological constant $\Lambda$ have faced significant challenges. Crucially, it was recognized that the short-distance structure of the RG-improved spacetime is determined by the running of $\Lambda$~\cite{Koch:2013owa}, which reinstates the  curvature singularity of the classical Schwarzschild solution. This highlights  that deforming a Schwarzschild-de Sitter (SdS) classical solution simply by replacing in this solution the bare coupling constants -- $G_{0}$ and $\Lambda_{0}$ -- with their running counterparts, can lead to inconsistencies in the deep Planckian region. 

In fact, constructing the quantum-improved geometry requires a diffeomorphism-invariant identification of the scale $k$ in terms of the radial coordinate $r$. However, this identification of $k(r)$ is not independent of the spacetime metric; it necessarily depends on the quantum-corrected (or dressed) metric itself. As a result, directly substituting $G_{k}$ and $\Lambda_{k}$ into the classical SdS solution is not fully consistent, especially in the ultraviolet (UV) regime, where significant deviations from the classical solution are expected. The feed-back between the running couplings and the metric must be accounted for.  This issue has only been partially addressed in subsequent studies~\cite{Platania:2019kyx}. Another limitation of earlier derivations of Refs.~\cite{Bonanno:2000ep,Koch:2013owa} is the challenge of obtaining a unique, global form for the function $k(r)$ that smoothly interpolates between the UV and IR regimes. This leads to some arbitrariness in the global metric solution describing the whole quantum-deformed spacetime.

In this work, we revisit the original derivation or Ref.~\cite{Bonanno:2000ep} by implementing a self-consistent identification of the scale $k$ in terms of the radial coordinate $r$. Our goal is to construct the effective spacetime geometry by self-consistently deforming the classical SdS solution using the running of the gravitational couplings within a FRG framework. This is achieved by regarding the identification of~\cite{Bonanno:2000ep} as a differential equation for $k(r)$, which tracks the evolution of $G_{k}$ and $\Lambda_{k}$ along the RG flow in a self consistent manner. Our procedure not only ensure full self-consistency, but also leads to a unique determination of the function $k(r)$. By numerically integrating the differential equation for $k(r)$ employing a shooting method we ware  able to derive the global form of quantum-deformed black hole metric.

The key outcome of our investigations is the discovery of a much richer UV phase structure in the quantum-improved geometry than previously known. This complex UV structure features a conical singularity, both Anti de Sitter (AdS) and dS phases, and a mass threshold for the formation of horizons. 

Our findings also align with recent studies that explores the UV critical manifold beyond polynomial truncations of the Einstein-Hilbert Lagrangian. These investigations, which primarily focused on a conformally-reduced theory of the form \( g_{\mu\nu} = \phi^2 \delta_{\mu\nu} \), have uncovered the existence of a \emph{pre-geometric} phase of unbroken diffeomorphism invariance at super-Planckian energies. This phase is characterized by the vanishing of the mean value of the conformal factor \(\phi\):
\begin{equation*}
    \langle \phi \rangle = 0\, .
\end{equation*}
If such a phase persists in the complete theory, it would suggest a fundamental breakdown of the classical notion of spacetime at extreme high energies~\cite{Reuter:2008qx,Daum:2008gr,Bonanno:2012dg,Bonanno:2023fij}. This would imply that the RG improvement approach may no longer hold at ultra-Planckian scales.

The structure of this paper is the following. In \cref{sect:eqg}, we provide a brief overview of the main tools required for the constructing effective quantum geometries. In \cref{sect:uvas} and \cref{sect:iras}, we use the Frobenius method to derive the approximate form of the quantum geometry in the UV and IR regimes, respectively. In \cref{sect:is}, we employ a shooting method to numerically construct the solution that interpolates between the UV and the IR approximate solutions.
In \cref{sect:euvslrgf}, we discuss the case in which the RG trajectories near the UV fixed point are not given by an approximate analytic form, but instead by a linear approximation. \Cref{sect:csad} is devoted to the computation of anomalous dimension using the conformal sector. Finally, in \cref{sect:c}, we present our conclusions.

\section{Effective quantum geometries }
\label{sect:eqg}
At classical level, the gravitational interaction is described by the Einstein-Hilbert (EH) action with a cosmological constant $\Lambda$.  
In the FRG framework of the Asymptotic-Safety program, quantum corrections are included by integrating quantum fluctuations up to the momentum scale $k^2$. The renormalized dynamic is encoded in an EAA $\Gamma_{k}[g_{\mu\nu}]$, where $g_{\mu\nu}$ represents the spacetime metric. Generally, $\Gamma_{k}[g_{\mu\nu}]$ is a solution to the FRG equations~\cite{Bonanno:2000ep,Koch:2013owa,Platania:2019kyx}, which in general cases cannot be solved analitically.  
However, owing to the fact that the UV critical manifold is finite-dimensional, only a finite number of relevant operator are required to control the continuum limit. 
Thus, when investigating quantum black holes, we can reasonably  restrict ourselves to the EH truncation, where the gravitational part of the EAA is given by \footnote{Throughout the paper, we adopt natural units $\hbar= c = 1$, with the Planck mass denoted by $\mathit{m}_{p}$ and the Planck length by $\ell_{p}$.}
\begin{equation}\label{EHT}
    \Gamma_{k}[g_{\mu\nu}] = \frac{1}{16 \pi G_{k}} \int d^{4} x \sqrt{-g} [R - 2\Lambda_{k}] \, ,
\end{equation}
and includes two scale-dependent couplings: the running Newton constant $G_{k}$ and the running cosmological constant $\Lambda_{k}$. The beta functions resulting from the EH truncation are expressed in terms of the dimensionless coupling constants
\begin{equation}
	g(k) = G_{k} k^{2} \quad \lambda(k) = \Lambda_{k} k^{-2} \label{ScalingRelation} \, ,
\end{equation}
whose scale dependence is better understood as a flow in the 2-dimensional subspace spanned by $g(k)$ and $\lambda(k)$~\cite{Koch:2013owa}.
The validity of the EH truncation at every momentum scale $k$ is quite a strong assumption, especially near the NGFP, where higher-derivative terms could become important. Nonetheless, the EH truncation \eqref{EHT} represents the leading term in the derivative expansion of $\Gamma_{k}[g_{\mu\nu}]$ and is expected to provide the leading contribution to the quantum-deformed black-hole geometry.

We consider static, spherically-symmetric solutions of the theory. At the classical level, these solutions are described by the well-known SdS solution 
\begin{equation}\label{gs}
	\dd s^{2} = -f(r) \dd t^{2} + \frac{\dd r^{2}}{f(r)} + r^{2} \dd\Omega^{2}\, ,
\end{equation}
where $\dd\Omega^{2} = \dd\theta^{2} + \sin^{2}\theta \dd\phi^{2}$ and 
\begin{equation}{\label{ClassicalBHSolution}}
	f(r) = 1 -  \frac{R_\text{s}}{r} - \frac{\Lambda_0 r^{2}}{3} \, .
\end{equation}
Here, $R_\text{s} = 2 G_{0} M$ is the Schwarzschild radius, while $G_0$ and $\Lambda_0$ represent the IR values of the couplings, i.e., the observed Newton constant and cosmological constant, respectively.
The present observed value of $\Lambda_0$ is quite small and it can be set to zero whenever black holes are considered. Thus, throughout this paper, we assume
\begin{equation}
\label{L0}
	\Lambda_0 = 0\,.
\end{equation}
The classical solution \eqref{ClassicalBHSolution} is modified by quantum gravity corrections encoded in the running of the coupling constants. There are several possibilities to take this running into consideration. The most general approach is to account for the running  at the level of the EAA \eqref{EHT}. A second, intermediate, option is to include the running of $g(k)$ and $\lambda(k)$ directly at the level of the field equations. Finally, the simplest possibility is to introduce the running directly at level of the solution~\cite{Bonanno:2000ep,Koch:2013owa}.  
In this paper, we adopt the latter approach, assuming that the quantum dressed solutions retains the same form as \eqref{gs}, but with the metric function modified by the running, 
\begin{equation}{\label{MetricFunctionDressed}}
	f_{\text{d}}(r) = 1- \frac{2 G_{k} M}{r} - \frac{\Lambda_{k} r^{2}}{3} \, .
\end{equation}
While this assumption is again quite strong, it is expected to capture the leading quantum deformations~\cite{Bonanno:2020bil}.
Notice that we are implicitly assuming that quantum gravity effects can be effectively described by a smooth geometry, which may not hold in the $k\to \infty$ UV regime, where the classical GR description is expected to break down. However, the hope is that the theory will naturally reveal its own limits of validity. We will revisit this point in the next section when discussing the effective UV geometries.

Similarly to Ref.~\cite{Koch:2013owa} but in contrast to Refs.~\cite{Bonanno:2000ep,Platania:2019kyx}, we are assuming that there is a contribution to the dressed black-hole solution stemming from the running of the cosmological constant $\lambda(k)$. This implies that while the dS term in \cref{MetricFunctionDressed} vanishes in the IR due to \cref{L0}, it still plays a pivotal role in determining the quantum-dressed solution. As we will see later, this is particularly significant in the UV regime, where the running cosmological constant impacts the structure of the solution.

\subsection{Renormalization-Group trajectories}
To determine the quantum-dressed black-hole solutions, we first need the form of the running couplings $g(k)$ and $\lambda(k)$. These couplings are obtained by solving the RG equations, i.e., by determining the RG trajectories once the beta functions $\beta_{g(k)},\beta_{\lambda(k)}$ are calculated using the EH truncation \eqref{EHT} (see, e.g., Ref.~\cite{Koch:2013owa}).
The phase diagram of the resulting autonomous ordinary differential equation (ODE) system has been constructed in Ref.~\cite{Koch:2013owa}. It is governed by a UV NGFP located at $\lambda_*=0.193$, $g_*= 0.707$ which acts as an UV attractor (and providing the UV completion of the theory) and a Gaussian fixed point (GFP) in the IR. Depending on the value of $\Lambda_0$, the flow exhibit a crossover in the IR~\cite{Koch:2013owa}.   
For our purposes, we first use the approximated analytic form of the RG trajectories provided in Ref.~\cite{Koch:2013owa}
\begin{widetext}
    \begin{subequations}{\label{RGFlowEquationApprox}}
\begin{align}
	g(k) & = \frac{G_{0} k^{2}}{1+\frac{G_{0}}{g_{*}} k^{2}} \, , \\
	\lambda(k) & = \frac{g_{*}\lambda_{*}}{g(k)} \left\{ \left( 5 + \frac{\Lambda_{0} G_{0}}{g_{*}\lambda_{*}} \right)[1-g(k)/g_{*}]^{3/2} - 5 + \frac{3g(k)}{2g_{*}}\left( 5 - \frac{g(k)}{g_{*}} \right) \right\} \, .
\end{align} 
\end{subequations}
\end{widetext}
In \cref{sect:euvslrgf}, we will also consider a linear approximate solution near the NGFP. The value of the parameters in the UV and IR, $\lambda_*$, $g_*$, $G_0$, $\Lambda_0$, fully determine the RG trajectory. In the following, we will set $\Lambda_0=0$ in accordance with \cref{L0}, while $G_0$ is identified with the Newton constant.

\subsection{Identification of the momentum scale in terms of spacetime coordinates}
Given its dependence on the RG scale $k$, \cref{MetricFunctionDressed} does not describe a well-defined spacetime geometry. To properly describe the geometry we must convert the momentum scale $k$ into a position dependent quantity. To do this, we first notice that the coarse-graining momentum scale must be invariant under coordinate transformations. This requirement is satisfied by constructing $k(P)$ in terms of the proper distance $\mathscr{L}(P)$ of a point $P$ in the black-hole spacetime
\begin{equation}{\label{CutoffIdentification}}
	k(P) = \frac{\xi}{\mathscr{L}(P)} \, ,
\end{equation}
where $\xi$ is a dimensionless, order-one parameter that will be fixed by imposing physical conditions.
Following the arguments in Ref.~\cite{Bonanno:2000ep,Koch:2013owa,Platania:2019kyx}, this identification must respect the symmetries of the classical solution. 
Therefore, we choose $\mathscr{L}(P)$ as the proper distance along a purely radial curve $\gamma_r$
\begin{equation}{\label{GeodesicEquation}}
	\mathscr{L} (r) = \int_{\gamma_r} \dd s =\int \frac{\dd r}{\sqrt{f_{\text{d}}(r)}} \, .
\end{equation}
This establishes a functional relationship between the coarse-graining momentum scale and the radial coordinate, giving $k=k(r)$.
However, this expression is not easy to work with. The problem is that the RG-scale $k$ depends on the form of the dressed metric function \eqref{MetricFunctionDressed}, which in turns depends -- through the form of $g(k)$ and $\lambda(k)$ given by \cref{RGFlowEquationApprox} -- on the identification of $k$ in terms of $r$ given by \cref{CutoffIdentification,GeodesicEquation}.
This sort of feed-back between dressed spacetime geometry and the form of $k(r)$ makes the problem of determining $f_{\text{d}}(r)$ quite involved. Until now, the identification \eqref{CutoffIdentification} and \eqref{GeodesicEquation} has been employed either neglecting the feed-back completely, i.e., using the classical, undressed, form of the metric function $f(r)$ into \cref{GeodesicEquation}~\cite{Bonanno:2000ep,Koch:2013owa}, or using an iteration procedure (a self-adjusting cutoff which depends on the initial, bare, background metric)~\cite{Platania:2019kyx}. In both cases, however, the procedure lacks full self-consistency. As a result, the solutions obtained are rough approximations, rather than exact representations of the quantum-dressed black-hole geometry.

In this paper, we will fully account for the feed-back between the dressed spacetime geometry and the RG scale $k$ to determine the exact form of the dressed metric function $f_{\text{d}}(r)$. Specifically, we will combine \cref{MetricFunctionDressed,CutoffIdentification,RGFlowEquationApprox} to write $f_{\text{d}}(g(k),\lambda(k),r)$ as a function of $\mathscr{L}(r)$. Then, we will reformulate \cref{GeodesicEquation} as a first order, nonlinear ODE for $\mathscr{L}(r)$, as follows
\begin{equation}{\label{GeodesicEquation1}}
	\mathscr{L}^{\prime} (r) = \frac{1}{\sqrt{f_{\text{d}}(\mathscr{L},r)}} \, ,
\end{equation}
where $f_{\text{d}}(\mathscr{L},r)=f_{\text{d}}(g(\xi/\mathscr{L}),\lambda(\xi/\mathscr{L}),r)$ is obtained by using $g(k)$, $\lambda(k)$, given by \cref{RGFlowEquationApprox}, into  \cref{MetricFunctionDressed}. We have 
\begin{widetext}
    \begin{equation}{\label{DimensionedMetricFunction}}
\begin{split}
    f_{\text{d}}(r) &= 1 - \frac{R_\text{s}}{r \left(1+\frac{G_{0} \xi ^2}{g_{*} \mathscr{L}(r)^2}\right)}-\frac{g_{*} \lambda_{*}  r^2}{3 G_{0}} \left(\frac{G_{0} \xi ^2}{g_{*} \mathscr{L}(r)^2}+1\right) \times \\
    &\times \left[-5 + 5 \left(1-\frac{G_{0} \xi ^2}{g_{*} \mathscr{L}(r)^2 \left(\frac{G_{0} \xi ^2}{g_{*} \mathscr{L}(r)^2}+1\right)}\right)^{3/2}+\frac{3 G_{0} \xi ^2 \left(5-\frac{G_{0} \xi ^2}{g_{*} \mathscr{L}(r)^2 \left(\frac{G_{0} \xi ^2}{g_{*} \mathscr{L}(r)^2}+1\right)}\right)}{2 g_{*} \mathscr{L}(r)^2 \left(\frac{G_{0} \xi ^2}{g_{*} \mathscr{L}(r)^2}+1\right)}\right] \, ,
\end{split}
\end{equation}
\end{widetext}
After solving \cref{GeodesicEquation1}, the function $\mathscr{L}(r)$ is recovered, which, once inserted in \cref{DimensionedMetricFunction}, provides the exact solution for the dressed metric function $f_{\text{d}}(r)$. 
In the next two sections, we will tackle this ODE by using  a series expansion near the UV fixed point and in the IR. This approach will allow us to determine the asymptotic behavior of the quantum-dressed geometry in both limits. Following these analytic results, we will also compute in the next sections the exact interpolating solution by numerically integrating \cref{GeodesicEquation1}.

\section{UV approximate solutions}{\label{UVApproximateSolutionsSection}}
\label{sect:uvas}

UV approximate solutions of \cref{GeodesicEquation1} can be obtained using the Frobenius method. We begin by expanding the function $k(r)^2$ as a series near $r=0$:
\begin{equation}
    k(r)^2=\sum_{n=-N}^{\infty} a_n  r^n \, ,
\end{equation}
where $N$ is a positive integer. The coefficients $a_n$ and the index $N$ are determined by substituting this ansatz into \cref{GeodesicEquation1} and applying the standard recurrence method. The leading $r=0$ behavior of the proper distance $\mathscr{L}(r)\sim r^{3/2}$ in the classical Schwarzschild geometry~\cite{Bonanno:2000ep,Koch:2013owa} suggests choosing $N=3$. However,  this choice is not self-consistent, as substituting it into \cref{GeodesicEquation1} leads to an imaginary value for the parameter $\xi$.  
Thus, accounting for the geometry's feed-back in \cref{GeodesicEquation} makes the choice $\mathscr{L}(r)\sim r^{3/2}$ of Ref.~\cite{Bonanno:2000ep,Koch:2013owa} inconsistent. We are, therefore, led to set $N=2$ into the near $r=0$ expansion \footnote{This choice also aligns with the na\"ive interpretation of $\mathscr{L}$ as the radius of the transverse $2-$sphere.}.
As we will see, certain features of the UV geometries derived in Refs.~\cite{Koch:2013owa,Bonanno:2000ep} arise as artifacts of using the wrong exponent $N$. 

To investigate the implications of using the right exponent $N$, we would like to check whether features such as the excision of the curvature singularity at $r=0$, as predicted in Ref.~\cite{Koch:2013owa}, is still realized. Other features, like the dS behavior of the UV geometry~\cite{Koch:2013owa,Bonanno:2000ep} will survive, but only in an approximate sense.

With $N=2$, the near $r=0$ expansion reads
\begin{equation}\label{kUV}
	k^{2} (r) = \frac{\beta}{r^{2}} + \frac{\gamma}{r} + \delta + \mathcal{O}(r) \, ,
\end{equation}
where $\beta$, $\gamma$ and $\delta$ are coefficients to be determined.
The leading term in the near $r=0$ expansion of the proper distance \eqref{CutoffIdentification} is linear 
\begin{equation}\label{LPD}
	\mathscr{L} = c r+\mathcal{O}(r^2) \, .
\end{equation}
Inserting  \cref{kUV} into \cref{GeodesicEquation1} and requiring it to hold up to order $r^3$, we find
\begin{subequations}
\begin{align}\label{kUVc}
	\xi & =\frac{1}{\sqrt{\lambda_{*}}} \frac{\sqrt{3\sigma}}{\sqrt{3-\sigma}}\, ; \\
	\gamma & = \frac{6 g_{*} M}{\sigma -6}\, ; \\
	\delta & = \frac{g_{*} \lambda_{*}}{2 \sigma (2 \sigma -9)}\left[\frac{18 g_{*} M^2 (33-7 \sigma )}{(\sigma -6)^2}-\frac{7 \sigma^{2} }{G_{0} \lambda_{*}}\right]\, ,
\end{align}
\end{subequations}
where we have defined $\beta \lambda_{*}= \sigma$ for simplicity. In this parametrization, $\sigma$ is the sole free parameter. Requiring the reality of $\xi$ and the absence of divergences in $\xi$ and $\delta$ constrain $\sigma$ to the interval  
\begin{equation}{\label{BetaCondition}}
	0 < \sigma < 3 \, .
\end{equation}
Once $k(r)$ is determined up to linear order in $r$, we can easily obtain the form of the dressed metric function \eqref{MetricFunctionDressed} in the UV regime up to $\mathcal{O}\left(r^3\right)$:
\begin{equation}{\label{DressedMetricFunctionUV}}
f_{\text{d}}^{\text{(UV)}}(r)  = \left(1-\frac{\sigma }{3}\right)-\omega r- 
	\frac{r^2}{L^2}+\mathcal{O}\left(r^3\right) \, ,
\end{equation}
where
\begin{equation}\label{tp1}
\begin{split}
    \frac{1}{L^{2}} &= 9 \mathcal{M}^2\frac{(3-\mathit{\sigma})(24-5\mathit{\sigma})}{{\sigma}^2(\mathit{\sigma}-6)^2(9-2 \mathit{\sigma})} - \frac{7  \mathcal{M}^2_{p}}{2}\frac{3-\sigma}{9-2 \mathit{\sigma}}\, ; \\ 
    \omega &= \frac{4  \mathcal{M} (\sigma -3)}{\sigma  (\sigma -6)} \, ,
\end{split}
\end{equation}
where we have defined the black hole's scaled mass $\mathcal{M} = M g_{*}\lambda_{*}$ and scaled Planck mass $\mathcal{M}^{2}_{p} = g_{*}\lambda_{*} /G_{0} = m_{p}^{2} g_{*}\lambda_{*}$.
The metric function \eqref{DressedMetricFunctionUV} characterizes the quantum-dressed geometry in the UV regime. Depending on the sign of $1/L^2$, it represents either a dS or an AdS spacetime with a singularity at $r=0$. The latter, however, is milder than the conventional Schwarzschild curvature singularity $R_\text{s}/r^3$. Specifically, the FRG flow of the cosmological constant in the UV generates a new conical singularity term (the $ \sigma/3$ term in \cref{DressedMetricFunctionUV}), responsible for a singular contribution $\sim \sigma /r^2$ to the scalar curvature. Moreover, the flow of $G_N$ accounts for a $\sim \omega /r$ term, stemming from the linear contribution in \cref{DressedMetricFunctionUV}.
This result differs from that of Ref.~\cite{Koch:2013owa}, where the flow in the UV interchanges the regular cosmological constant term $\sim \Lambda r^2$ with the singular Schwarzschild term $R_\text{s}/r$ in the metric function.

It is interesting to mention that the UV form of the metric \eqref{DressedMetricFunctionUV} has been obtained, in a different context, as an exact black-hole solution in conformal Weyl gravity~\cite{Mannheim} \footnote{In \cref{DressedMetricFunctionUV}, the Schwarzchild term is absent in the UV regime, but emerges in the IR regime, as discussed in the next section.} This similarity is not coincidental, as conformal symmetries are expected to hold at the UV NGFP. These symmetries are broken by the RG flow, generating the two singular terms in \cref{DressedMetricFunctionUV}. 
The conical singularity, induced by the flow in the UV of the cosmological constant term, entails the breakdown of our effective spacetime description in terms of a smooth geometry. Its presence signals the onset of a pre-geometric phase of unbroken diffemorphism invariance, uncovered in Ref.~\cite{Reuter:2008qx}.
 
In the following, we will remove the short-distance curvature singularities by cutting the spacetime. Although this will result in a geodesically-incomplete geometry, it provides valuable physical insights into the behavior at short distances.

\subsection{UV geometry}{\label{sect:uvg}}
To remove the curvature singularities from our spacetime, we begin by applying a translation $r\to r+\mathcal{L}$ to the radial coordinate. This allows to get rid of the linear term in the metric function \eqref{DressedMetricFunctionUV} yelding 
\begin{equation}\label{uvgeometry}
\begin{split}
    d\text{s}^{2}_{\text{d}} &= - f_{\text{d}}^{\text{(UV)}} dt^{2} + \frac{dr^{2}}{f_{\text{d}}^{\text{(UV)}}} + (r + \mathcal{L})^{2} d\Omega^{2}, \\ 
    f_{\text{d}}^{\text{(UV)}}(r) &= 1 -\mathcal{C}- \frac{r^{2}}{L^{2}} + \mathcal{O}(r^3)\, , 
\end{split}
\end{equation}
with 
\begin{equation}\label{tp}
	 \mathcal{L} = - \frac{L^2 \omega}{2}\, , \qquad \mathcal{C}=\frac{\sigma }{3}-\frac{L^2 \omega^2}{4}\, .
\end{equation}
With the new radial coordinate, the conical singularity is moved to $r= - \mathcal{L}$, which is positive (negative) for $L^2$ negative (positive).

The form  of the UV geometry crucially depends on the sign of $L^2$, which determines the effective cosmological constant  $\Lambda_{\text{eff}}=3/L^2$. Quite interestingly, the  mass gives a \emph{positive} contribution to $\Lambda_{\text{eff}}$, whereas the Planck mass contributions are always \emph{negative}. Physically, this competition determines a transition from a regime with $\Lambda_{\text{eff}} < 0$, controlled by $\mathit{m}_{p}$ and corresponding to a domination of the vacuum energy, to a regime with $\Lambda_{\text{eff}}>0$, in which matter excitations, controlled by $M$, becomes dominant.

The transition between these two regimes occurs at the critical value
\begin{equation}\label{tra}
		M_{t}^2 =\frac{1}{g_{*}\lambda_{*}}\frac{7}{18}
  \frac{(\mathit{\sigma}-6)^2 \mathit{\sigma}^2}{ 24 - 5\mathit{\sigma}} \mathit{m}^2_{p} \, . 
	\end{equation}
 In the following, we will discuss these two cases separately. First notice that, at the critical transition value $M=M_{t}$, the UV geometry corresponds to a Minkowski spacetime with a conical singularity, which can be removed by cutting the spacetime at finite geodesic length.

\subsubsection{Short distance  $\text{AdS}_2\times \text{S}^2$ geometry}

For $M<M_{t}$ it follows from \cref{tp,BetaCondition} that $1/L^2$ is negative, while $\mathcal{L}$ remains positive. 
To remove the conical singularity, we cut the spacetime at $r=0$. In the vicinity of $r=0$, the metric \eqref{uvgeometry} describes a regular four-dimensional spacetime where the geometry factorizes as $\text{AdS}_2\times \text{S}^2$, with the radius of the two-sphere given by $\mathcal{L}$. Notice that $\mathcal{C}$ in 
\cref{uvgeometry} can simply be removed through a coordinate rescaling of the $\text{AdS}_2$ spacetime, setting $r\to  c t$, $t\to t/c$, $c^2=1-\mathcal{C}$. 
Under this transformation, the singularity shifts to a negative value of the new radial coordinate $r=-\mathcal{L} $ placing it  outside the range of positive $r$. While this procedure removes the singularity from the physical region, the spacetime is still singular since we are cutting it at finite geodesic length.

\subsubsection{Sub-Plankian objects}

For sub-Planckian objects with $M \ll m_{p}$, \cref{uvgeometry} describes a compact object with a short-distance, $\text{AdS}_2\times \text{S}^2$ geometry.
In this regime, the first term in \cref{tp1} can be neglected and the effective UV cosmological constant is negative, fully determined by the Planck mass, and reads
	\begin{equation}
	{\Lambda_{\text{eff}}^{\text{(UV)}}}\sim -\frac{21 \mathcal{M}^2_{p}}{2}\frac{\sigma-3}{2 \mathit{\sigma}-9 }\, .
	\end{equation}
Starting from the $\text{AdS}_2\times \text{S}^2$, sub-Planckian object and increasing the mass, we have Planck-scale objects for which both $L$ and the radius of the two-sphere are of the order of the Planck length.  
  
\subsubsection{Short distance  $\text{dS}_2\times \text{S}^2$ geometry} 
For $M>M_{t}$, the effective cosmological constant is positive, while $\mathcal{L}$ turns negative. Therefore, the conical singularity is now located at positive values of $r=|\mathcal{L}|$.
To avoid the latter, we cut the spacetime at $r=-2\mathcal{L}$. 
Near this point, the spacetime geometry behaves like   $\text{dS}_2 \times \text{S}^2$. 
Similarly to the previous case, removing the curvature singularity results in a geodesically-incomplete spacetime.  

\subsubsection{Phase transition in the UV geometry}

A striking result of our derivation is the presence of a phase transition in the UV geometry at the critical mass value \eqref{tra}, which is of order of $\mathit{m}_{p}$. Geometrically, this transition marks  the change in the UV geometry shifting from $\text{AdS}_2\times \text{S}^2$ to  $\text{dS}_2\times \text{S}^2$ .

This phase transition in gravity was first predicted using general and field-theoretical arguments by Polyakov~\cite{Polyakov:1993tp}. The scale dependence of the effective gravitational coupling in the IR $G_k/G_0=1 +\kappa (G_0 k^2)$ along with the existence of a continuum limit implies  a phase transition at a mass scale $M_{t}$ of order $G_0^{-1/2}$, i.e., $M_{t}\sim  \mathit{m}_{p}$.
Moreover, the existence of an $\text{AdS}_2\times \text{S}^2$ phase in quantum gravity, in particular in connection with the near-horizon geometry of extremal black holes, has already emerged in different approaches~\cite{Cadoni:2023tse}. Furthermore, $\text{AdS}_2$ quantum gravity exhitib several appealing features, making it quite useful for addressing challenging problems in black-hole physics, such as the information puzzle and the microscopic origin of black hole entropy~\cite{Maldacena:1998uz,Almheiri:2019qdq,Penington:2019kki,Cadoni:2023tse,Bonanno:2007wg}. Its appearance within the FRG approach to quantum gravity gives further support to its general relevance for quantum gravity. It is also possible that its emergence near a NGFP could be linked to the existence of a phase with unbroken conformal symmetries, a common  feature of both fixed points in the FRG approach and $\text{AdS}_2$ quantum gravity.

There is, however, an important distinction between the $\text{AdS}_2\times \text{S}^2$ phase associated with the near-horizon of near-extremal black holes and the short distance $\text{AdS}_2\times S_2$ phase found in this paper for FRG quantum gravity. In the former case, the geometry describes an infinite throat in the near-horizon near-extremal limit, whereas in the latter, the spacetime is truncated at a finite geodesic length.  

\subsubsection{Astrophysical black holes} 

For astrophysical objects ($M\gg \mathit{m}_{p}$), the effective UV cosmological constant is primarly determined by the mass and is given by
 \begin{equation}
\Lambda_{\text{eff}}^{\text{(UV)}}\sim 27 \mathcal{M}^2\frac{(3-\mathit{\sigma})(24-5\mathit{\sigma})}{{\sigma}^2(\mathit{\sigma}-6)^2(9-2 \mathit{\sigma})} \, .
 	\end{equation}
The approximate UV form of the metric \eqref{uvgeometry} does not allow us to distinguish between objects with and without horizons. In \cref{sect:is}, we will show that a threshold mass exists for the formation of event horizons.

For super-Planckian objects, particularly astrophysical black holes with masses significantly exceeding the critical threshold $M_{t}$, the behavior at length scales much larger then the Planck length resembles a black hole with a dS core. This result aligns with the findings of Refs.~\cite{Bonanno:2000ep, Koch:2013owa}, which both predict a dS geometry in the black-hole interior. In this regime feed-back effects of the geometry in \cref{GeodesicEquation1} are negligible.
Our predictions differ from those of Refs.~\cite{Bonanno:2000ep, Koch:2013owa} in the deep UV regime, where, instead, the feed-back effects become relevant. Notably, the appearance of a curvature singularity at $r=0$, predicted in~\cite{Koch:2013owa}, arises as an artifact originated by neglecting the feed-back. In our UV geometry, this curvature singularity is replaced first by a conical singularity and then by an $\text{AdS}_2 \times \text{S}^2$ geometry for sub-Planckian black holes, and by $\text{dS}_2 \times \text{S}^2$ for Planckian objects with $M>M_{t}$.

\section{IR approximate  solutions}
\label{sect:iras}

Let us now seek approximate solutions of \cref{GeodesicEquation1} in the $k\to 0$
regime, corresponding to the $r\to \infty$ expansion. Since we are assuming a vanishing IR value for the cosmological constant, the leading behavior of the metric function in the $r\to \infty$ expansion will be given by the standard Schwarzschild solution, i.e., by \cref{gs} with $\Lambda_0=0$.

To compute the dressed metric function \eqref{MetricFunctionDressed} in the IR regime, we will follow the same procedure as in the previous section. First, we will find an approximate series solution of \cref{GeodesicEquation1} and then compute the form of the dressed metric function in the IR, denoted as $f_{\text{d}}^{\text{(IR)}}(r)$.

We stress that, even though we assume a vanishing $\Lambda_0$, the IR dressed metric will still receive contributions from the running cosmological constant $\Lambda_k$ in \cref{MetricFunctionDressed}. As we will see below, this term will contribute subleading corrections to the classical $1/r$ Schwarzschild term.

As before, we use an  ansatz for the asymptotic form of the momentum scale $k(r)$, or equivalently, for the proper distance $\mathscr{L}(r)$. This ansatz can be easily inferred by considering the behavior in the classical Schwarzschild geometry~\cite{Bonanno:2000ep,Koch:2013owa}
\begin{equation}\label{hhh}
	\mathscr{L}(r) = r + \mathcal{O}\left[\ln\left(\sqrt{\frac{r}{R_\text{s}}}\right)\right] \, .
\end{equation}
We are therefore led to the asymptotic, $r \to \infty$, expansion 
\begin{widetext}
    \begin{equation}\label{LIR}
	\mathscr{L}(r) \sim \textsf{a} r + \textsf{b} \ln \left(\sqrt{\frac{r}{R_\text{s}}}\right)+ \textsf{d} + \textsf{e} \frac{ \ln \left(\sqrt{r/R_\text{s}}\right)}{r}+\frac{\textsf{f}}{r} + \frac{\textsf{h} \ln \left(\sqrt{r/R_\text{s}}\right)}{r^2}+\frac{\textsf{i}}{r^2} + \mathcal{O}\left[\frac{\ln(\sqrt{r/R_\text{s}})}{r^{3}}\right]  \, ,
\end{equation}
\end{widetext}
where, $\textsf{a}$, $\textsf{b}$, $\textsf{e}$, $\textsf{f}$, $\textsf{h}$, $\textsf{i}$ are constant coefficients. Inserting this equation into \cref{GeodesicEquation1} yields the following values of the coefficients:
\begin{subequations}\label{pir}
	\begin{align}
		\textsf{a} & = 1\, ; \\
		\textsf{b} & = R_\text{s}\, ; \\
		\textsf{e} & = 0 \, ;\\
		\textsf{f} & = -\frac{G_{0}\zeta ^4 \lambda_{*} +6 g_{*} R_\text{s}^2}{16 g_{*}}\, ; \\
		\textsf{h} &  = \frac{\zeta ^4 G_{0} \lambda_{*}  R_\text{s}}{8 g_{*}}\, ; \\
		\textsf{i} & = \frac{G_{0}R_\text{s} \zeta ^2  (8-\frac{1}{2}\zeta ^2 \lambda_{*})+4 \textsf{d} \zeta ^4 G_{0} \lambda_{*}-5 g_{*} R_\text{s}^3 }{32 g_{*}}\,.
	\end{align}
\end{subequations}
$\textsf{d}$, instead, plays the role of an integration constant in the differential equation \eqref{GeodesicEquation1}.

Using these results into \cref{MetricFunctionDressed} yields the dressed metric function in the IR approximation. Up to order $\mathcal{O}\left[\ln^{2}(\sqrt{r/R_\text{s}})/r^4\right]$, it reads as
\begin{widetext}
    \begin{equation}{\label{IRApproximatedMetricFunction}}
	f_{\text{d}}^{\text{(IR)}}(r) = 1-\frac{R_\text{s}}{r} - \frac{\zeta ^4 G_{0} \lambda_{*} }{8 g_{*} r^2 }  + \frac{2 \zeta ^2 G_{0} R_\text{s} +\textsf{d} \zeta ^4 G_{0} \lambda_{*} +R_\text{s} \zeta ^4 G_{0} \lambda_{*}  \ln \left(\sqrt{r/R_\text{s}} \right)}{2  g_{*} r^3} + \mathcal{O}\left[\frac{\ln^{2} \left(\sqrt{r/R_\text{s}}\right) }{r^{4}}\right]\, .
\end{equation}
\end{widetext}
There are still two free parameters, $\textsf{d}$ and $\zeta$. The latter is determined by requiring consistency with the UV expansion, leaving $\textsf{d}$ as the only free parameter.

The subleading terms relative to the classical $1/r$ term represent long-range, IR corrections induced by quantum effects. The coefficients of the $1/r^2$, $\ln \left(r/R_\text{s}\right)/r^3$ and $1/r^3$ terms can thus be interpreted as  quantum hair for the classical black-hole solution. Apart from the term proportional to $\textsf{d}$, these contribution are naturally of Planckian origin and suppressed by the Planck mass. The only one which could be super-Planckian is $\textsf{d}$. It is noteworthy that the $1/r^2$, $\ln r/r^3$ terms and $1/r^3$ terms arise from the flow of $\lambda(k)$ in the IR, since they are proportional to $\lambda_{*}$. This explains why these terms are absent in the IR expression given in Ref.~\cite{Bonanno:2000ep}, where only the $1/r^3$ term proportional to $G_{0} R_\text{s}$, is present.

The $1/r^2$ term originates from the the flow of the $r^2/L^2$ UV term in the IR. It may be viewed as the long-range gravitational contribution of vacuum polarization by conformal fields degrees of freedom at the UV fixed point. This interpretation finds support  from the calculation  the energy density of the (anisotropic) fluid sourcing the gravitational field
\begin{equation}
\rho_v= -\frac{\zeta^4}{32 \pi}\frac{\lambda_*}{g_*}\frac{1}{r^4}\, ,
\end{equation}
which is the typical (negative) vacuum energy density expected for conformal fields in $4$D~\cite{Cadoni:2023nrm}.

Regarding the $\sim 1/r^3$ and $\ln \left(r/R_\text{s}\right)/r^3$ contributions, there are two types of terms: $1)$ those scaling as $G_0 R_\text{s}$ and $2)$ those scaling as $G_0 \textsf{d}$.  The former arise from the flow of the UV term $r^2/L^2$ into the IR. Consequently, the associated quantum hair $\ell$ follows the general scaling law $\ell \sim L^{2/3}R_\text{s}^{1/3}$ (see Ref.~\cite{Cadoni:2022chn}). Indeed with $L\sim \ell_{p}$, we find $\ell^3\sim G_0^2 M$, which correctly yields the terms of the form $\sim \ell^3/r^3$.  

In general, these terms describe mass polarization effects. The energy densities of the effective (anisotropic) fluid sourcing them in the dressed metric are given by
\begin{equation}
\begin{split}
\rho_1 &= \frac{\zeta^2}{4  g_* \pi}\frac{R_\text{s}}{r^5}\, ,\\
\rho_2 &= \frac{R_\text{s}\lambda_*\zeta^4}{8  g_* \pi}\left(\frac{\ln \left(\sqrt{r/R_\text{s}}\right)}{r^5}+\frac{1}{r^5}\right)\, ,
\end{split}
\end{equation}
corresponding to the first and second $1/r^3$ terms in \cref{IRApproximatedMetricFunction}, respectively.

The second type of terms, being proportional to the arbitrary integration constant $\textsf{d}$, are not necessarily fixed by the flow from the UV to the IR. Although a naturalness argument would suggest $\textsf{d}\sim R_\text{s}$, it remains possible for $\textsf{d}$ to be determined purely by IR physics. For instance, it could be related to the presence of non-local terms in the IR EA \eqref{EHT}, derived in the $k\to 0 $ limit~\cite{Belgacem:2017cqo}. In such a case, the quantum hair $\ell=\textsf{d}$ could also become super-Planckian (see for instance~\cite{Cadoni:2022chn}). 

In the remainder of this paper, we will not pursue this further possibility and will assume that such non-local effects are absent, consistently setting $\textsf{d}\sim R_\text{s}$.

In the next section we will numerically compute the global interpolating solution using a shooting method, with $\textsf{d}$ serving as the shooting parameter. We shall see that any choice of $\textsf{d}$ hierarchically different from $R_\text{s}$ will lead to naturalness issues in the shooting procedure.  

\section{Interpolating solutions}
\label{sect:is}
In the previous section, we derived the quantum dressed solution near $r=0$ and in the $r\to \infty$ region through  asymptotic expansions. Our next task is to determine the form of the solution in the whole domain $r\in[0,\infty)$ interpolating between these two asymptotic solutions. The quite intricate form of the differential equation \eqref{GeodesicEquation1} precludes any straightforward analytical approach. Therefore, we will employ a shooting method to numerically solve it.

\subsection{Shooting method}{\label{ShootingMethod}}

The shooting method consists in a numerical procedure in which we solve the same differential equation in two different regions of the domain. In the present case, in the near $r=0$ core and in the $r\to \infty$ asymptotic region. These solutions are then matched by defining an appropriate junction point. To ensure that the  interpolating solution is continuous and differentiable across the entire domain, one must vary the free parameters of the local solutions until the Wronskian of the two solutions, calculated at the junction point, cancels.

The first step is to rewrite the differential equation \eqref{GeodesicEquation} in terms of dimensionless variables. The equation contains two physical length scales, $\ell_{p}$ and $R_\text{s}$. Given that the classical solution is expressed solely in terms of $R_\text{s}$, we will use it to define dimensionless variables:
\begin{equation}\label{rv}
    \rho = \frac{r}{R_\text{s}}, \, \quad \mathcal{D}(\rho) = \frac{\mathscr{L}(\rho)}{R_\text{s}} \, .
\end{equation}
Written in terms of these new variables, \cref{GeodesicEquation} becomes
\begin{equation}{\label{AdimensionalisedGeodesicEquation}}
    \frac{\dd \mathcal{D}(\rho)}{\dd\rho} = \frac{1}{\sqrt{f_{\text{d}}(\rho)}} \, ,
\end{equation}
with $f_\text{d}(\rho)$ given by \cref{DimensionedMetricFunction}. It is also natural to measure the physical parameters $G_0$ and $M$ in units of $R_\text{s}$. Consequently, we define the positive definite dimensionless parameters, $\mathcal{G}$ and $\mu$
\begin{equation}{\label{AdimensionalParameters}}
    \mathcal{G} = \frac{G_{0}}{R_\text{s}^{2}} =  \frac{m_{p}^{2}}{4 M^2}\, ,\quad \mu =\mathcal{M}R_\text{s}= 2 g_{*} \lambda_{*} \frac{M^{2}}{m_{p}^{2}}\, .
\end{equation}
These two parameters are not independent, but are related by an inverse proportionality
\begin{equation}{\label{RelationMuG}}
    \mu = \frac{g_{*}\lambda_{*}}{2}\frac{1}{\mathcal{G}} \, .
\end{equation}
To implement the shooting method, we must numerically solve \cref{AdimensionalisedGeodesicEquation} separately for $\rho \ll 1$ (UV regime) and for $\rho \gg 1$ (IR regime) and then join the resulting solutions at a junction point. 
Let $\mathcal{D}_{\text{UV}}(\rho)$ and $\mathcal{D}_{\text{IR}}(\rho)$ represent the solutions in the UV and IR regimes, respectively. The Wronskian of these solutions is given by
\begin{equation}{\label{Wronskian}}
    W(\mathcal{D}_{\text{UV}},\mathcal{D}_{\text{IR}}) = \frac{\dd \mathcal{D}_{\text{UV}}(\rho)}{\dd\rho} \mathcal{D}_{\text{IR}}(\rho) - \frac{\dd \mathcal{D}_{\text{IR}}(\rho)}{\dd\rho} \mathcal{D}_{\text{UV}}(\rho) \, .
\end{equation}
To numerically solve \cref{AdimensionalisedGeodesicEquation} in the two regions, we need to specify the boundary conditions for $\mathcal{D}$, denoted as $\mathcal{D}_{\text{0,UV}}(\rho_{\text{UV}})$, $\mathcal{D}_{\text{0,IR}}(\rho_{\text{IR}})$ at choosen points in the UV $\rho_{\text{UV}}$ and IR $\rho_{\text{IR}}$ regimes, respectively. The simplest choice is to make use of the results from the previous two sections, where we derived the approximate expressions of $\mathscr{L}$ in both the UV and IR. Thus, we set
\begin{equation}
    \mathcal{D}_{\text{UV,approx}}(\rho_{\text{UV}}) = \mathcal{D}_{\text{0,UV}}(\rho_{\text{UV}}) \, , \label{UVBoundaryCondition}
\end{equation}
\begin{equation}
    \mathcal{D}_{\text{IR,approx}}(\rho_{\text{IR}}) = \mathcal{D}_{\text{0,IR}}(\rho_{\text{IR}}) \, , \label{IRBoundaryCondition}
\end{equation}
where $\mathcal{D}_{\text{UV,approx}}$ and $\mathcal{D}_{\text{IR,approx}}$ can be easily obtained using \cref{kUVc,kUV,LIR,pir} along with the definitions \eqref{rv} and \eqref{AdimensionalParameters}. For $ \mathcal{D}_{\text{UV,approx}}$, we have
\begin{equation}{\label{ExpansionAdimensionalisedProperDistanceUV}}
   \mathcal{D}_{\text{UV,approx}}(\rho) \simeq  \frac{1.73 \rho  \sqrt{\sigma }}{\sqrt{3-\sigma } \sqrt{\sigma }}-\frac{0.35 \rho ^2 \sqrt{\sigma }}{\mathcal{G} \sqrt{3-\sigma } \sigma ^{3/2} (\sigma -6)} + \mathcal{O}(\rho^{3}) \, ,
\end{equation}
whereas ${D}_{\text{IR,approx}}$ reads as
\begin{widetext}
    \begin{equation}\label{for1}
\begin{split}
    \mathcal{D}_{\text{IR,approx}}(\rho)  \simeq & \,  \frac{\chi }{2}+\rho + \log \left(\sqrt{\rho }\right)+\frac{0.09\left(-\frac{46.63 \mathcal{G} \sigma ^2}{(3-\sigma )^2}-4.24\right)}{\rho } + \frac{8.24 \mathcal{G} \sigma ^2 \log \left(\sqrt{\rho }\right)}{\rho ^2 (3-\sigma )^2} \\
    &+ \frac{0.04 \left(\frac{93.26 \chi  \mathcal{G} \sigma ^2}{(3-\sigma )^2}-\frac{23.32 \mathcal{G} \sigma ^2}{(3-\sigma )^2}+ \frac{124.35 \mathcal{G} \sigma }{3-\sigma }-3.53\right)}{\rho ^2} \, ,
\end{split}
\end{equation}
\end{widetext}
where we have used the numerical values $\lambda_*=0.193$, $g_*= 0.707$. Furthermore, following our naturalness assumption for the integration constant $\textsf{d}$, we have set in \cref{LIR} $\textsf{d}=\chi R_\text{s} /2$ with $\chi$ a dimensionless, order-$1$ parameter.
A close inspection of \cref{AdimensionalisedGeodesicEquation} reveals that the numerical integration procedure can only proceed up to the points where the metric function $f_{\text{d}}(\rho)$ vanishes, as the differential equation becomes singular there. These points correspond to event horizons  of the spacetime geometry. Consequently, our shooting algorithm cannot  directly generate black-hole solutions.
Nevertheless, we can still  infer the existence  of black holes by starting the shooting procedure with parameter values that allow for horizonless solutions. We then gradually adjust the parameters until we  encounter an extremal black hole, which will be  followed by a configuration with two horizons.

Let us now describe  the shooting procedure. We have two independent shooting parameters $\mathcal{G}$ and $\chi$, which correspond physically to the black-hole mass and the IR quantum $1/r^3$ hair, respectively. The parameter $\sigma$ is not physically relevant. It is related to the free parameter $\xi$ appearing in the cutoff identification \eqref{CutoffIdentification} (see \cref{kUVc}). Therefore, the parameter $\sigma$ is held fixed during the shooting procedure. The qualitative behavior of the solution will be independent from it. We fix $\mathcal{G}$ in the UV and use the shooting procedure to determine the value of $\chi$ in the IR.  
The UV solution, defined by the boundary conditions \eqref{UVBoundaryCondition} at $\rho_{\text{UV}}$ and the fixed value of $\mathcal{G}$, is evolved towards the IR and matched with continuity at the point $\rho_{\text{J}} \in [0,\infty)$ with the IR solution, specified by the boundary conditions \eqref{IRBoundaryCondition} at $\rho_{\text{IR}}$. This matching determines a value $\chi_{0}$ for $\chi$.

To implement the computational procedure, we must first choose the boundary values $\rho_{\text{UV}}$ and $\rho_{\text{IR}}$ as well as the junction point $\rho_{\text{J}}$. Certain computational criteria must be met. To avoid computational problems at the boundary point $\rho = 0$, we choose instead a small, but still non-vanishing, UV boundary value $\rho_{\text{UV}} = 10^{-9}$. 
The junction point is set to $\rho_{\text{J}} = 50$, far enough to minimize any dependence on the UV boundary. For the IR boundary, $\rho_{\text{IR}}$ is chosen in such a way to ensure that, when varying $\rho$ within the interval $\rho \in [\rho_{\text{IR}}-1000,\,\rho_{\text{IR}}]$, the resulting value of $\chi_{0}$ from the shooting procedure changes by less than $1\%$, which will be therefore taken as the associated error. The chosen value is $\rho_{\text{IR}} = 46000$. Additionally, it has been verified that, once the UV and IR boundaries have been fixed, varying $\rho_{\text{J}}$, $\chi_{0}$ changes by less than $1\%$.
Finally, we have lso verified that, by varying the range of values of $\chi$ in which the Wronskian cancels, $\chi_{0}$ varies less than $1\%$.

To find $\chi_{0}$, we generate a grid of $1000$ equally-spaced points along the dimensionless radial coordinate $\rho$. Then, for selected values of the parameters $\sigma$ and $\mathcal G$ we calculate the Wronskian at each grid point. The value of $\chi_{0}$ that cancels the Wronskian is then obtained by interpolating the curve where the Wronskian is a function of $\chi$. We refer to \cref{tab:etaresultsvaryingG,tab:etaresultsvaryingsigma} for some examples of determination of $\chi_{0}$. Finally, we verified that the grid spacing provides the necessary resolution. We have checked that, if we increase the number of points in the list, $\chi_{0}$ varies by less than $1\%$.

\begin{figure}
    \centering
    \includegraphics[width=\linewidth]{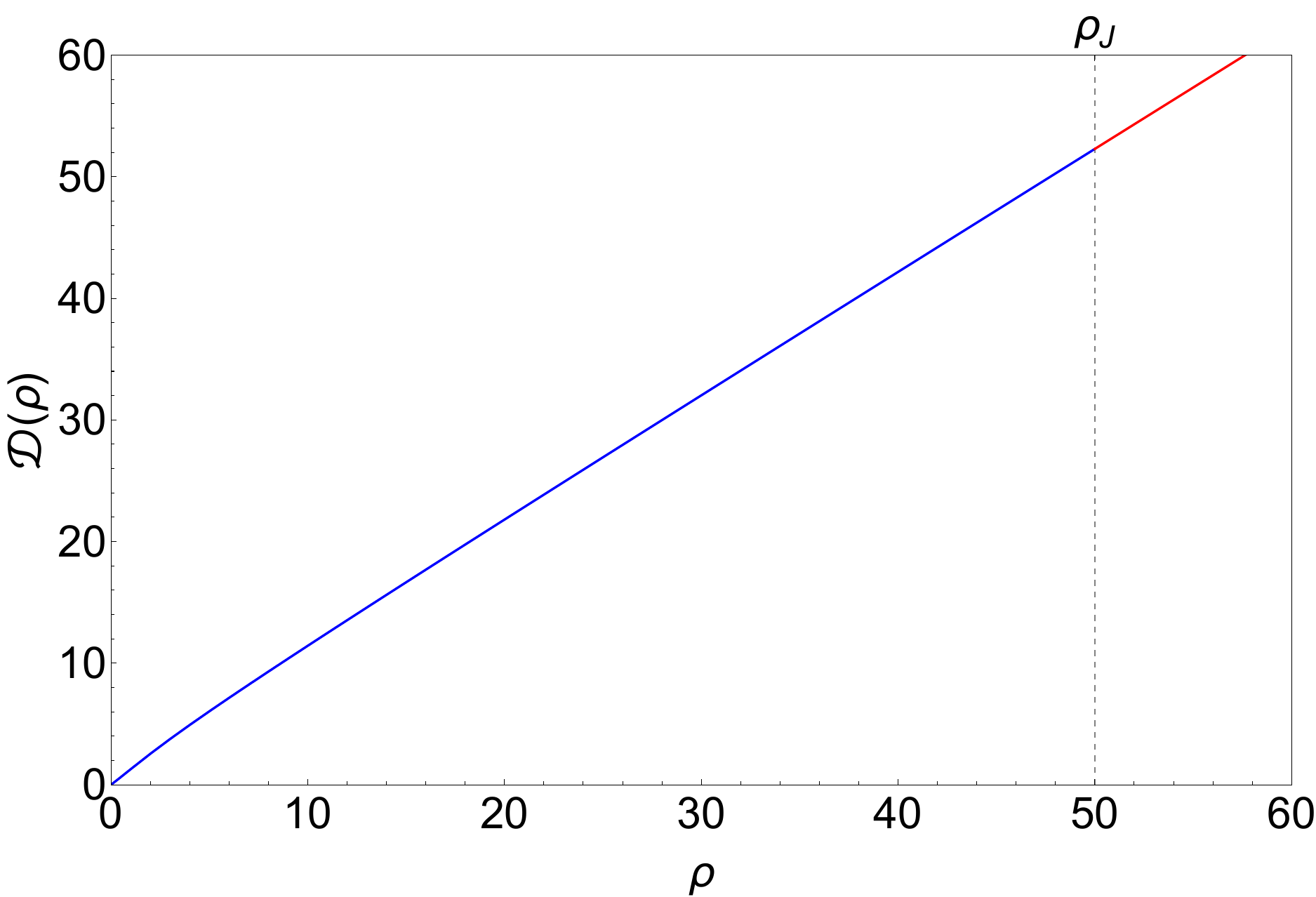}
    \includegraphics[width=\linewidth]{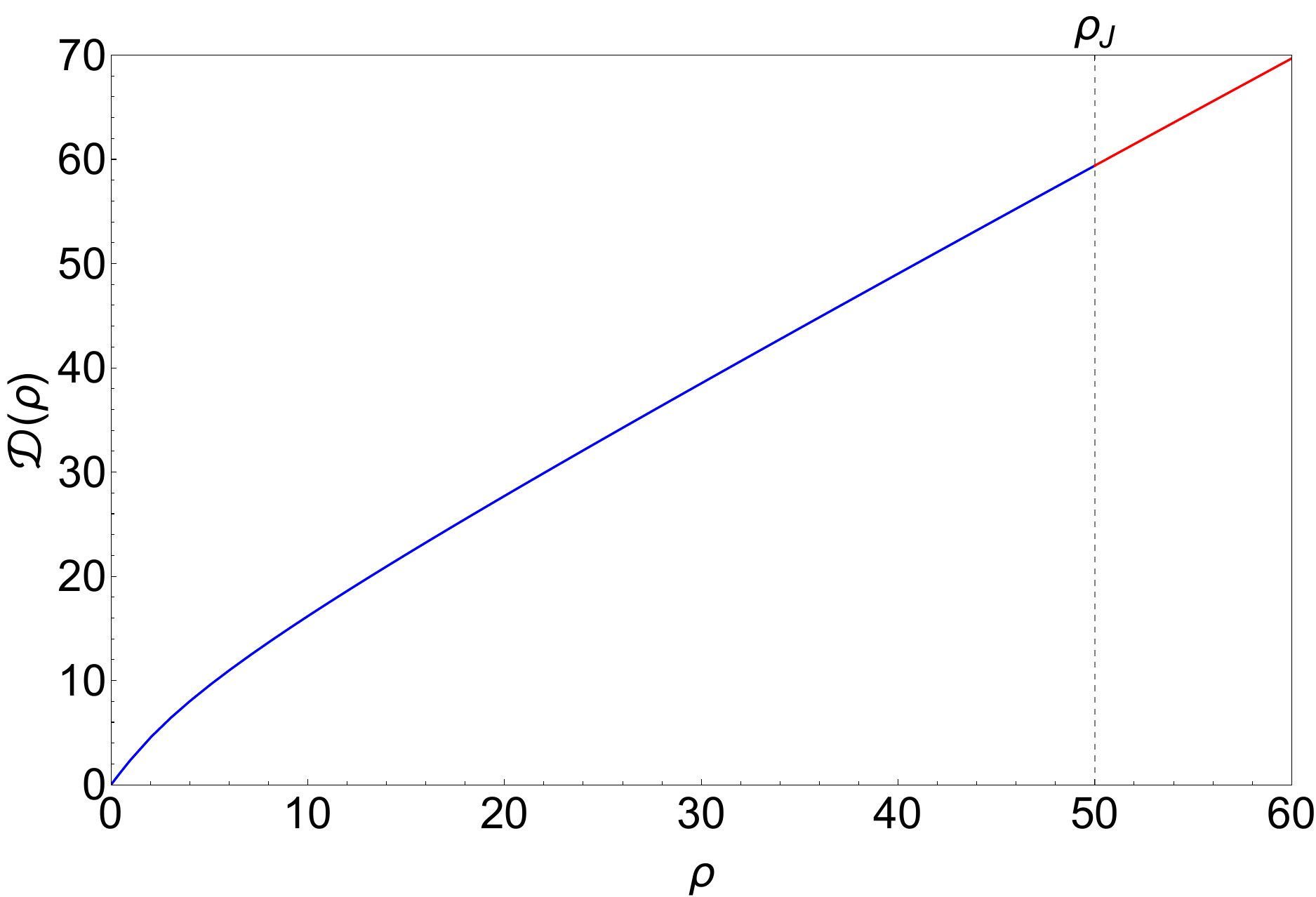}
    \caption{Numerical solutions of \cref{AdimensionalisedGeodesicEquation} are presented. The solutions are obtained by joining with continuity the UV solution with boundary condition \eqref{UVBoundaryCondition} (blue curve) with the IR solution with boundary condition \eqref{IRBoundaryCondition} (red curve) at the point $\rho_{\text{J}} = 50$. In the top panel, we show the solution for the parameter values $\sigma = 1$ and $\mathcal{G} = 0.5$, yelding an IR hair value of $\chi_{0} = 0.70\pm 0.01$. The bottom panel displays the solution for the parameter values $\sigma = 2.5$ and $\mathcal{G} = 1$, resulting in an IR hair value of $\chi_{0} = 17.82\pm 0.01$.}
    
    \label{fig:Adimensionalised_Proper_Lenght_Shooting}
\end{figure}

All the numerical curves for $\mathcal{D}$ obtained using the shooting method, including those shown in \cref{fig:Adimensionalised_Proper_Lenght_Shooting}, exhibit always the same qualitative behavior. They behave linearly both near $\rho=0$ and at large values of $\rho$,  with deviations from linearity occurring at intermediate values of the radial coordinate. This behavior is fully consistent with both the UV and IR analytic expansions \eqref{LPD} and \eqref{hhh}. Additionally, the deviations from linearity are minimal, as can also be seen through a visual inspection of \cref{fig:Adimensionalised_Proper_Lenght_Shooting}.

\begin{table}
    \centering
    \begin{tabular}{c|c|c|c|c|c}
    \toprule
\midrule
$\rho_{\text{UV}}$ & $\rho_{\text{J}}$ & $\rho_{\text{IR}}$ & $\sigma$ & $\mathcal{G}$ & $\chi_{0}$ \\
\hline
$10^{-9}$ & $50$ & $46000$ & $ 1$ & 0.2 & $1.32\pm 0.01$ \\
\hline
$10^{-9}$ & $50$ & $46000$ & $ 1$ & 0.5 & $0.70\pm 0.01$ \\
\hline
$10^{-9}$ & $50$ & $46000$ & $ 1$ & 1 & $0.56\pm 0.01$ \\
\hline
$10^{-9}$ & $50$ & $46000$ & $ 1$ & 7 & $1.50\pm 0.01$ \\
\midrule
\bottomrule
    \end{tabular}
    \caption{Determination of $\chi_0$ for a fixed value of the parameter $\sigma= 1$ and varying $\mathcal G$. The points $\rho_{\text{UV}}$, $\rho_{\text{J}}$ and $\rho_{\text{IR}}$ represents the values of the coordinate $\rho$ at the UV boundary, the junction and IR boundary, respectively.}
    \label{tab:etaresultsvaryingG}
\end{table}

\begin{table}
    \centering
    \begin{tabular}{c|c|c|c|c|c}
    \toprule
\midrule
$\rho_{\text{UV}}$ & $\rho_{\text{J}}$ & $\rho_{\text{IR}}$ & $\sigma$ & $\mathcal{G}$ & $\chi_{0}$ \\
\hline
$10^{-9}$ & $50$ & $46000$ & $ 0.5$ & 1 & $0.20\pm 0.01$ \\
\hline
$10^{-9}$ & $50$ & $46000$ & $ 1$ & 1 & $0.56\pm 0.01$ \\
\hline
$10^{-9}$ & $50$ & $46000$ & $ 1.5$ & 1 & $2.06\pm 0.01$ \\
\hline
$10^{-9}$ & $50$ & $46000$ & $ 2.5$ & 1 & $17.82\pm 0.01$ \\
\midrule
\bottomrule
    \end{tabular}
    \caption{Determination of $\chi_0$ for a fixed value of the parameter ${\mathcal G}=1$ and varying $\sigma$. The points $\rho_{\text{UV}}$, $\rho_{\text{J}}$ and $\rho_{\text{IR}}$ represents the values of the coordinate $\rho$ at the UV boundary, the junction and IR boundary, respectively.}
    \label{tab:etaresultsvaryingsigma}
\end{table}

\subsection{Numerical determination of the metric function}
\begin{figure}
    \centering
    \includegraphics[width=\linewidth]{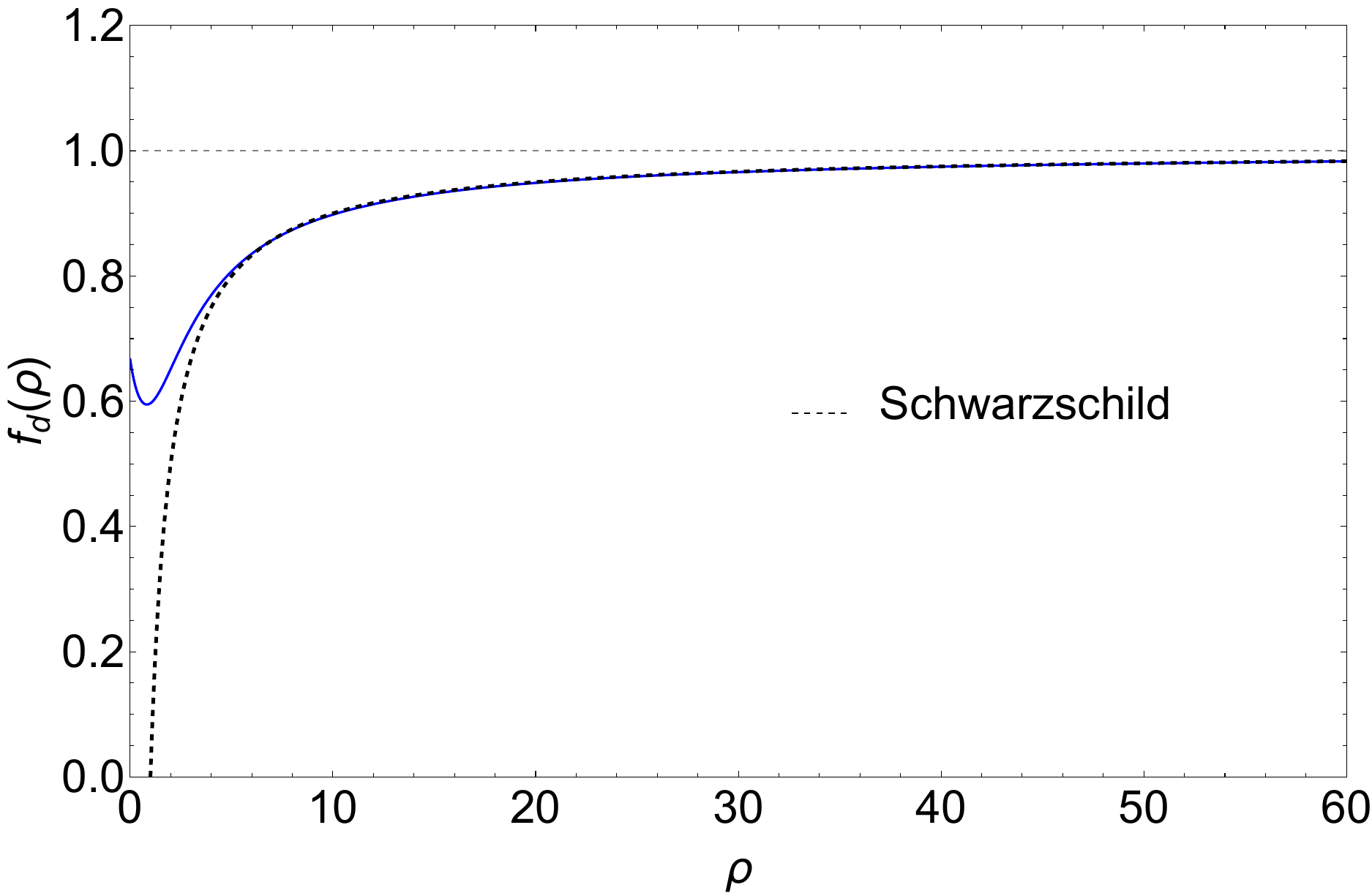}
    \includegraphics[width=\linewidth]{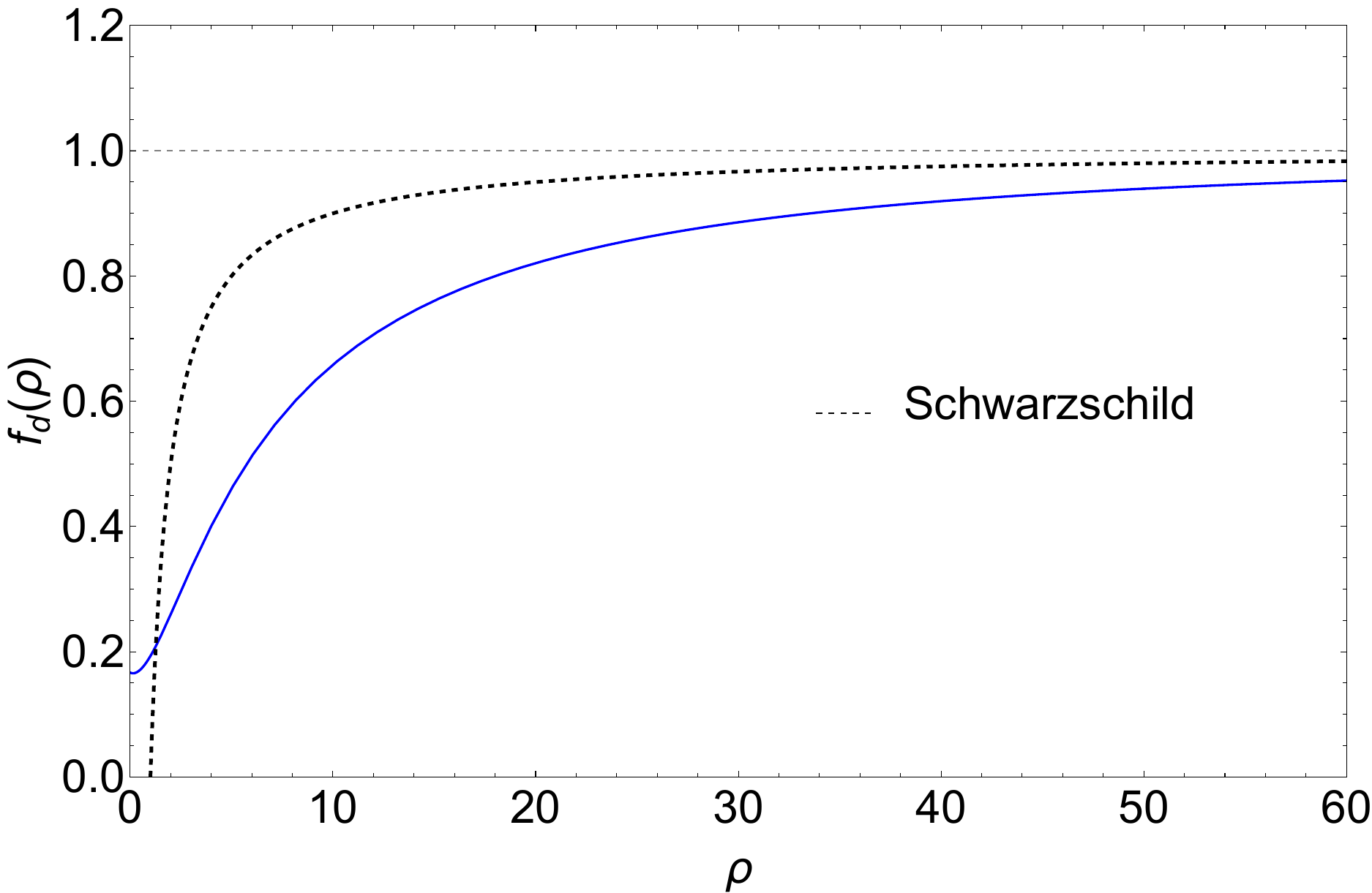}
    \caption{Comparison of the global behavior of the quantum-dressed metric function $f_{\text{d}}(\rho)$ with the classical Schwarzschild metric function for selected values of the parameters $\mathcal{G}$ and $\sigma$. Upper panel: plot of the curves for  $\sigma = 1$ and $\mathcal{G} = 0.5$ (corresponding to $\chi_{0} = 0.70\pm 0.01$). Bottom panel: plot of the curves with $\sigma = 2.5$ and $\mathcal{G} = 1$, (corresponding to $\chi_{0} = 17.82\pm 0.01$). }
    \label{fig:Space_Time_Shooting}
\end{figure}

Once the numerical solutions for $\mathcal{D}$ in \cref{AdimensionalisedGeodesicEquation} are obtained, we can determine the numerical form of the dressed metric function using \cref{DimensionedMetricFunction}.
In \cref{fig:Space_Time_Shooting} we present these results  for selected values of the parameters $\sigma$ and $\mathcal{G}$, comparing them with the classical Schwarzschild solution. 
The metric function is shown using \cref{DressedMetricFunctionUV}, which differs from the other form of the UV metric function \eqref{uvgeometry} by a translation of the radial coordinate $r\to r+\mathcal{L}$.

In the region outside the classical horizon $\rho>1$, the quantum-dressed solution closely resembles the Schwarzschild metric function $f_\text{Schw}(r)$. In the the top panel of \cref{fig:Space_Time_Shooting}, we show the dressed metric function $f_{\text{d}}(\rho)$ for a black hole with mass $M=\mathit{m}_{p}/\sqrt{2}$ (${\mathcal{G}}=1/2$). The plot reveals that $f_{\text{d}}(\rho)$ overlaps the classical metric function after few Schwarzschild radii ($\sim 5$). This becomes even more evident when $M> \mathit{m}_{p}$. However, $f_{\text{d}}(\rho)$ and $f_\text{Schw}(r)$ exhibit significant differences in the core region $\rho<1$. While the classical metric function decreases and intersects the horizontal axis at $\rho=1$, where the classical horizon forms, $f_{\text{d}}(\rho)$ instead features a minimum near $\rho=1$. At $\rho =0$ one would expect a maximum with $f_{\text{d}}(\rho=0)=1$, corresponding to a pure dS UV behavior. However, the presence of the conical singularity term $\mathcal{C}$ in \cref{uvgeometry} prevents the curve from reaching this maximum resulting in $f_{\text{d}}(\rho=0)<1$. 

\begin{figure}
    \centering
    \includegraphics[width=\linewidth]{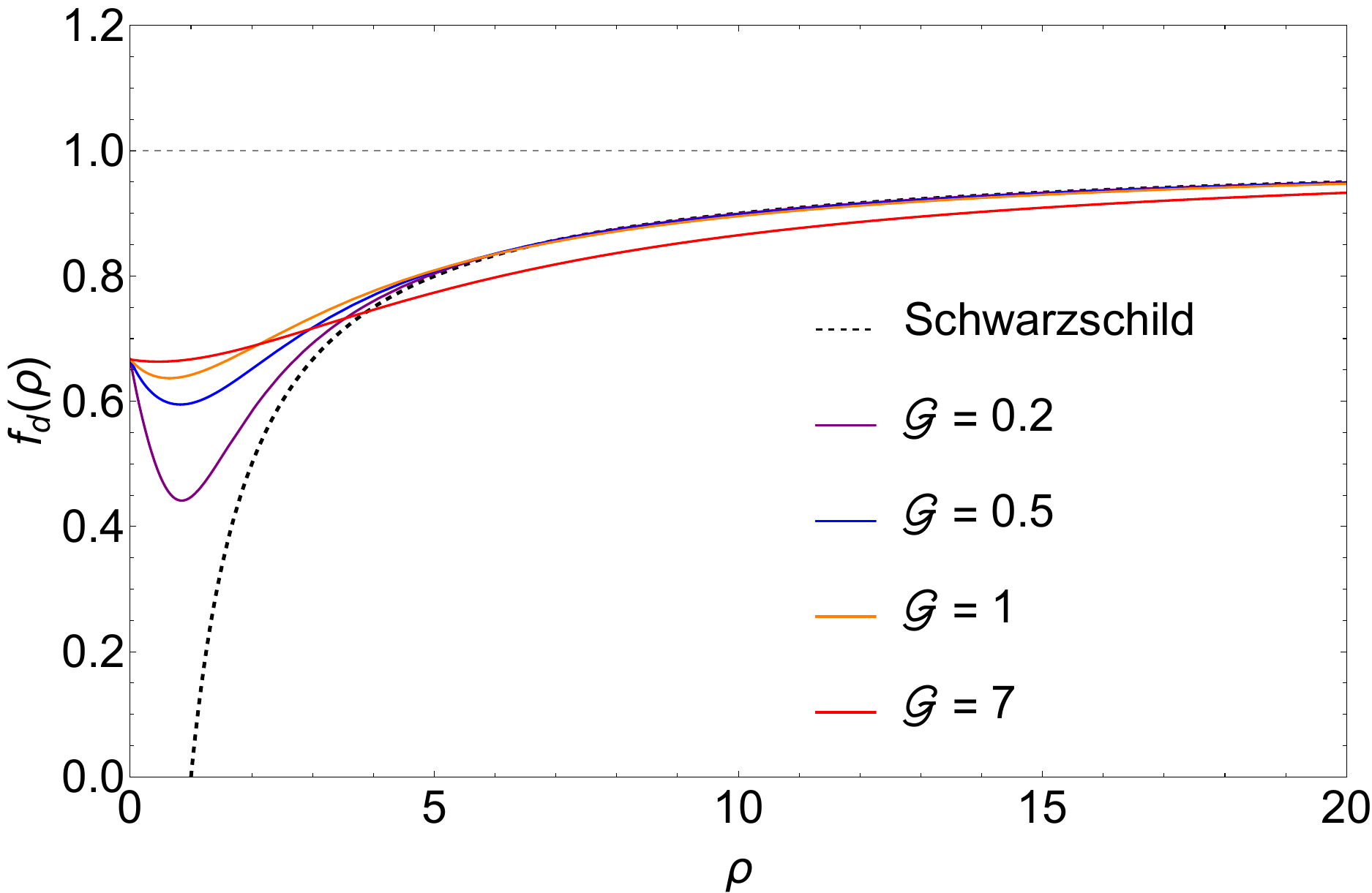}
    \caption{A zoomed-in view of the core region of the dressed metric function $f_{\text{d}}(\rho)$, showing the $\text{AdS}_2\times S^2$ $\to$ $\text{dS}_2\times S^2$ transition. The parameter $\mathcal{G}$ is varied from $\mathcal{G}=7$ to $\mathcal{G}=0.2$, while keeping $\sigma = 1$. For comparison, the dimensionless classical Schwarzschild metric function is also shown (dashed curve).}
    \label{fig:Detail_On_Space_Time_Core}
\end{figure}

For strictly sub-Planckian objects, such as the one shown in the bottom panel of \cref{fig:Space_Time_Shooting}, corresponding to $M=\mathit{m}_{p}/2$, the quantum-dressed object deviates more significantly from the classical black hole even in the region outside the classical horizon. Furthermore, in the core region $\rho<1$, the behavior of $f_{\text{d}}(\rho)$ is qualitative different from that found for $M\sim \mathit{m}_{p}$. One can easily see from the bottom panel in \cref{fig:Space_Time_Shooting} that now, in the core region, $f_{\text{d}}(\rho)$ reaches a minimum at $\rho=0$, corresponding to the $\text{AdS}_2\times \text{S}^2$ behavior in the core discussed earlier. 
By lowering the value of the parameter $\mathcal{G}$, i.e. by increasing the mass $M$, the second derivative changes sign and the concavity of the curve near the $\rho=0$ region changes, as can be easily inferred from \cref{fig:Detail_On_Space_Time_Core}. This change reflects the sign flip of the effective cosmological constant $\Lambda_{\text{eff}}$, which occurs as the mass decrease (see \cref{tp}). This is the transition $\text{dS}_2\times S_2$ $\to$ $\text{AdS}_2\times S_2$ described in \cref{sect:uvg}  occurring at the critical value $M_{c}$ of the mass given in \cref{tra}.

By further increasing the mass towards values $M>\mathit{m}_{p}$, i.e., by decreasing the dimensionless parameter $\mathcal{G}$, the minimum of the curve $f_{\text{d}}(\rho)$ moves towards the $\rho-$axis (see \cref{fig:Detail_On_Space_Time_Core}).

It is therefore expected that above a second critical value $M_{c}$ of the mass an horizon forms. The solution gives in this case an extremal black hole. For $M>M_{c}$, the curves  intersects the $\rho$ axis at two points, corresponding to the formation of inner and outer horizons. This two-horizon black hole solutions are qualitatively similar to regular black hole solutions a with a dS core~\cite{Bonanno:2000ep}. Conversely, for $M < M_{c}$ the solutions describe horizonless singular objects.  
Although our numerical shooting algorithm cannot directly generate solutions with horizons because of the divergence in \cref{AdimensionalisedGeodesicEquation}, we will still provide a plot of a black hole solution by interpolating $f_{\text{d}}(\rho)$ in the region between the two horizons.

Let us now discuss in detail the two transitions $\text{dS}_2\times S^2$ $\to$ $\text{AdS}_2\times S^2$ and horizonless solution $\to$ black hole, occurring respectively at $M=M_{t}$ and $M=M_{c}$.

\subsection{$\text{AdS}_2\times S^2$ $\to$ $\text{dS}_2\times S^2$ transition}

In the discussion of the UV approximated solutions presented in \cref{sect:uvg}, the transition has been predicted to occur at the value of $M_{t}$ given by \cref{tra}. In our numerical solution, this transition manifests itself as a change in the sign of the second derivative (the concavity) of the function $f_{\text{d}}(\rho)$ in the $\rho\sim 0$ region, when $\mathcal{G}$ is decreased below the critical value $\mathcal{G}_{t}$. 
For $\rho\gg 1$, the concavity of the curve remains the same, regardless of the value of $\mathcal{G}$, implying the presence of an inflection point $\rho_{t}$ for the curve corresponding to the critical parameter $\mathcal{G}_{t}$. 

\begin{figure}
    \centering
    \includegraphics[width=\linewidth]{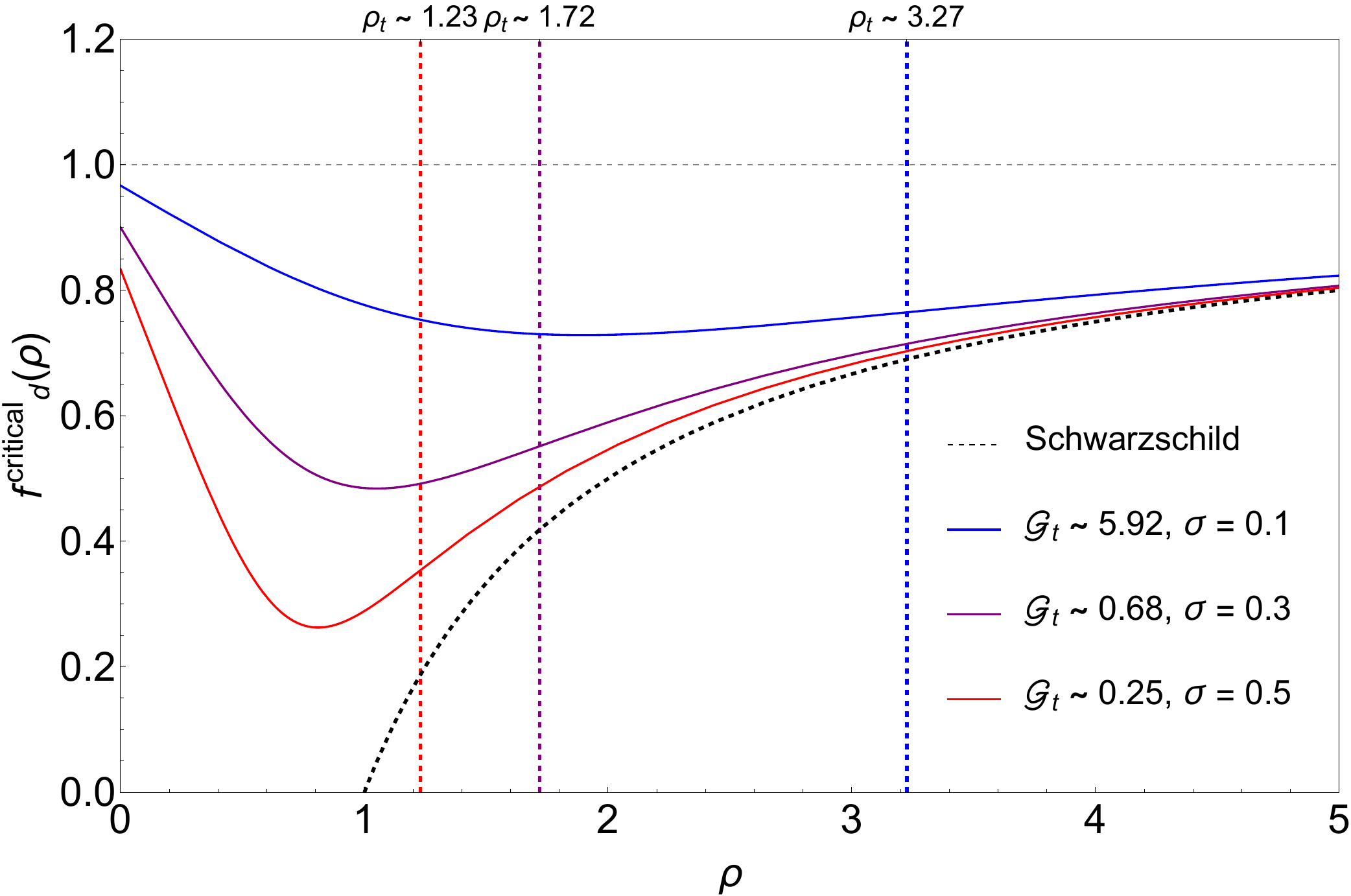}
    \caption{A comparison between the metric function $f_{\text{d}}^{\text{critical}}(\rho)$ at selected critical values of parameter $\mathcal{G}_{t}$ and that for selected values of 
    the parameter $\sigma$. The curves show a change in concavity at the inflection points $\rho_{t}$, which are indicated by vertical lines in the plot. For comparison, the dimensionless classical Schwarzschild metric function is also shown (dashed curve).}
    \label{fig:core_comparison}
\end{figure}

The change in the concavity of the curve $f_{\text{d}}(\rho)$ is analysed in \cref{fig:Detail_On_Space_Time_Core,fig:core_comparison}.
In \cref{fig:Detail_On_Space_Time_Core} we show the curves in the core region for solution obtained by increasing $\mathcal{G}$ from $\mathcal{G}=0.2$ to $\mathcal{G}=7$ at a selected fixed value of $\sigma = 1$. The change in concavity of the curve occurs between the orange curve ($\mathcal{G}=1)$ and the red curve ($\mathcal{G}=7)$.

In \cref{fig:core_comparison} we compare the critical curves $f_{\text{d}}^{\text{critical}}(\rho)$, corresponding to different critical values of $\mathcal{G}_{t}$ and $\sigma$. 
Vertical lines indicate the positions of the inflection points. Notably, as $ \sigma \to 0$, $f_{\text{d}}(\rho = 0) \to 1$, as the conical singulatity $\sigma/3$ in \cref{DressedMetricFunctionUV} vanishes. Additionally, as $ \sigma$ increases, the value of $\rho_{t}$ decreases and, causing, the minimum of the $f_{\text{d}}(\rho)$ curve to shift towards the $\rho=0$ direction. 
We have explicitly checked that the critical values of the parameter $\mathcal{G}_{t}$, calculated analytically using \cref{tra}, i.e., 
\begin{equation}{\label{GadimAtTransition}}
    \mathcal{G}_{t} = \frac{9}{14}\frac{(24-5 \sigma )}{ (\sigma -6)^2 \sigma ^2} g_{*} \lambda_{*}
\end{equation}
agree with those obtained through numerical integration. In \cref{tab:ValuesTransitiondSAdS}, we present a table with the numerical determination of the critical parameter $\mathcal{G}_{t}$ for the critical curves displayed in \cref{fig:core_comparison}, which match the analytical results from \cref{GadimAtTransition} within the  numerical error. 


\begin{table}
    \centering
    \begin{tabular}{c|c|c|c|c}
    \toprule
\midrule
$\rho_{\text{UV}}$ & $\rho_{\text{J}}$ & $\sigma$ & $\mathcal{G}_{t}$ & $\rho_{t}$ \\
\hline
$10^{-9}$ & $50$ & $ 0.1$ & $5.92\pm 0.01$ & $3.23\pm 0.01$ \\
\hline
$10^{-9}$ & $50$ & $ 0.3$ & $0.67\pm 0.01$ & $1.72\pm 0.01$\\
\hline
$10^{-9}$ & $50$ & $ 0.5$ & $0.25 \pm 0.01$ & $1.23 \pm 0.01$ \\
\midrule
\bottomrule
    \end{tabular}
    \caption{Numerical determination of the critical values of the parameter $\mathcal{G}_{t}$ for the critical curves presented in \cref{fig:core_comparison}. From left to right: $\rho_{\text{UV}}$ and $\rho_{\text{J}}$ represents the coordinates of UV boundary and the junction, respectively. $\mathcal{G}_{t}$ denotes the critical value of the parameter at the dS/AdS transition. Lastly, $\rho_{t}$ indicates the inflection point where the concavity of $f_{\text{d}}(\rho)$ changes.}
    \label{tab:ValuesTransitiondSAdS}
\end{table}

\subsection{Black-hole formation }
\label{subsec:BHform}

\begin{figure}
    \centering
    \includegraphics[width=\linewidth]{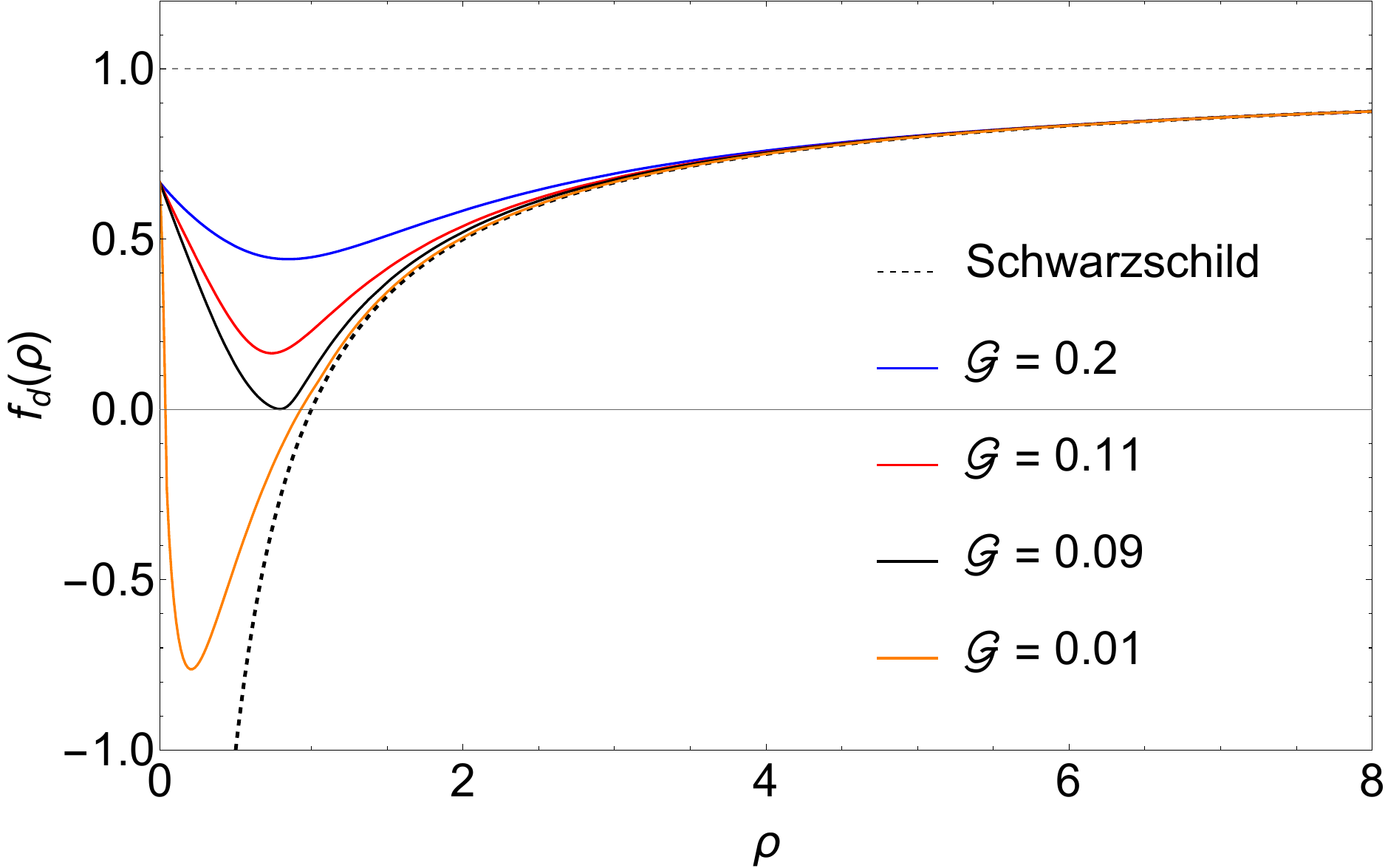}
    \caption{Horizon formation for solutions with $\sigma = 1$ in shown by plotting the metric function $f_{\text{d}}(\rho)$ for progressively decreasing values of the parameter $\mathcal{G}$. The extremal black hole forms for $\mathcal{G}= 0.09=\mathcal{G}_{c}$. The orange curve depicts a (non-extremal) black-hole solution featuring two horizons, with the region between the two horizons obtained using an interpolation algorithm. For comparison, the dimensionless classical Schwarzschild metric function is also shown (dashed curve).}
    \label{fig:CoreDetailsVaryingSigma}
\end{figure}

For every given value of the parameter $\sigma$ within the allowed range \eqref{BetaCondition}, increasing the mass above a critical value $M_{c}$ (or equivalently lowering $\mathcal{G}$ below a critical value $\mathcal{G}_{c}$) causes the curve for $f_{\text{d}}(\rho)$ to intersect the $\rho$ axis indicating the formation of horizons. $M_{c}$ is always of order of the Planck mass. The process of horizon formation is presented in detail in \cref{fig:CoreDetailsVaryingSigma} for fixed $\sigma = 1$.
Starting from $\mathcal{G}=0.2$ and lowering it, the minimum of the $f_{\text{d}}(\rho)$ curves drops and approaches the horizontal axis. 
Our numerical estimate for the critical value of the parameter $\mathcal{G}_{c}$ at which an extremal black hole forms is $\mathcal{G}_{c} = 0.09$ (we have checked that the metric function is zero at the degenerate horizon within two significant digits). This corresponds to the critical mass
\begin{equation}\label{tra1}
    M_{c} = (1.67 \pm 0.01) \, \mathit{m}_{p} \, .
\end{equation}
Increasing the mass beyond this value results in a solution that describes a black hole with two horizons. In  \cref{fig:CoreDetailsVaryingSigma}, we show the curve for $f_{\text{d}}(\rho)$ corresponding to a non-extremal black hole with $\mathcal{G}=0.01$.
The shooting method does not allow for direct computation of the curve in the region between the two horizons. The plot in this region has been therefore obtained by interpolating the solutions in the exterior regions.

\subsection{Regularity of the Event Horizons}
Everything discussed so far rests on the \cref{GeodesicEquation1}. The latter, however, becomes pathological at the  horizons, where $f_\text{d} = 0$, leading to a divergence in the derivative of the proper length. This is due to the event horizon being a coordinate singularity in Schwarzschild coordinates.
Although this singularity can be removed with an appropriate change of coordinates, we still need to check if this non-analyticity in the proper distance is reflected in the metric function through \cref{DimensionedMetricFunction}. Specifically, the regularity of the first and second derivative of the metric function, which in turn determine the black-hole surface gravity and the curvature invariants, may be compromised.

A coherent framework for describing quantum deformations of static and spherically-symmetric geometries, which addresses the aforementioned issues, was developed in Refs.~\cite{Sannino:2023fiw,Sannino:2024gvw}. The metric function describes a quantum-deformed Schwarzschild solution, whose near-horizon expansion is expressed as a power-law series of the proper distance measured from the horizon, to ensure a non-singular horizon. The coefficients of the expansion are fixed by requiring finite first and second derivatives of the metric function, thus ensuring finiteness of the Hawking temperature and curvature invariants.

However, the framework of Refs.~\cite{Sannino:2023fiw,Sannino:2024gvw} extensively relies on the  near-horizon local  behaviour of the metric function $f$. This makes the approach unsuitable for describing quantum deformations which are constructed interpolating the behaviour of the metric function in different spacetime regions. This is the case of our approach, where an interpolation of the proper distance between the infrared and ultraviolet regimes is performed (also as previously done by Bonanno and Reuter \cite{Bonanno:2000ep}). 

We begin with a preliminary analysis of the regularity of our solution close to the horizon. An inspection to \cref{GeodesicEquation1,DimensionedMetricFunction} reveals that the near-horizon behavior of our solutions cannot be a power-law series. Indeed, the usual leading behavior $\mathscr{L}(r) \sim \sqrt{r-r_{h}}$ (or, equivalently, $f_{d} \sim r-r_{h}$, typical of the vast majority of static and spherically-symmetric black holes, makes the system of equations \eqref{GeodesicEquation1} and \eqref{DimensionedMetricFunction} inconsistent.  This non-trivial behavior is primarily driven by the (A)dS term. Additionally, due to the feed-back between the dressed spacetime geometry and the RG scale $k$, reconstructing the exact functional form of the metric function near the event horizon from the numerical solution proves to be challenging.  In particular this implies, not only that the metric function $f$ is not analytic at the horizon but  also that  this non-analiticity  is quite non trivial.

\begin{figure}
    \centering
    \includegraphics[width=\linewidth]{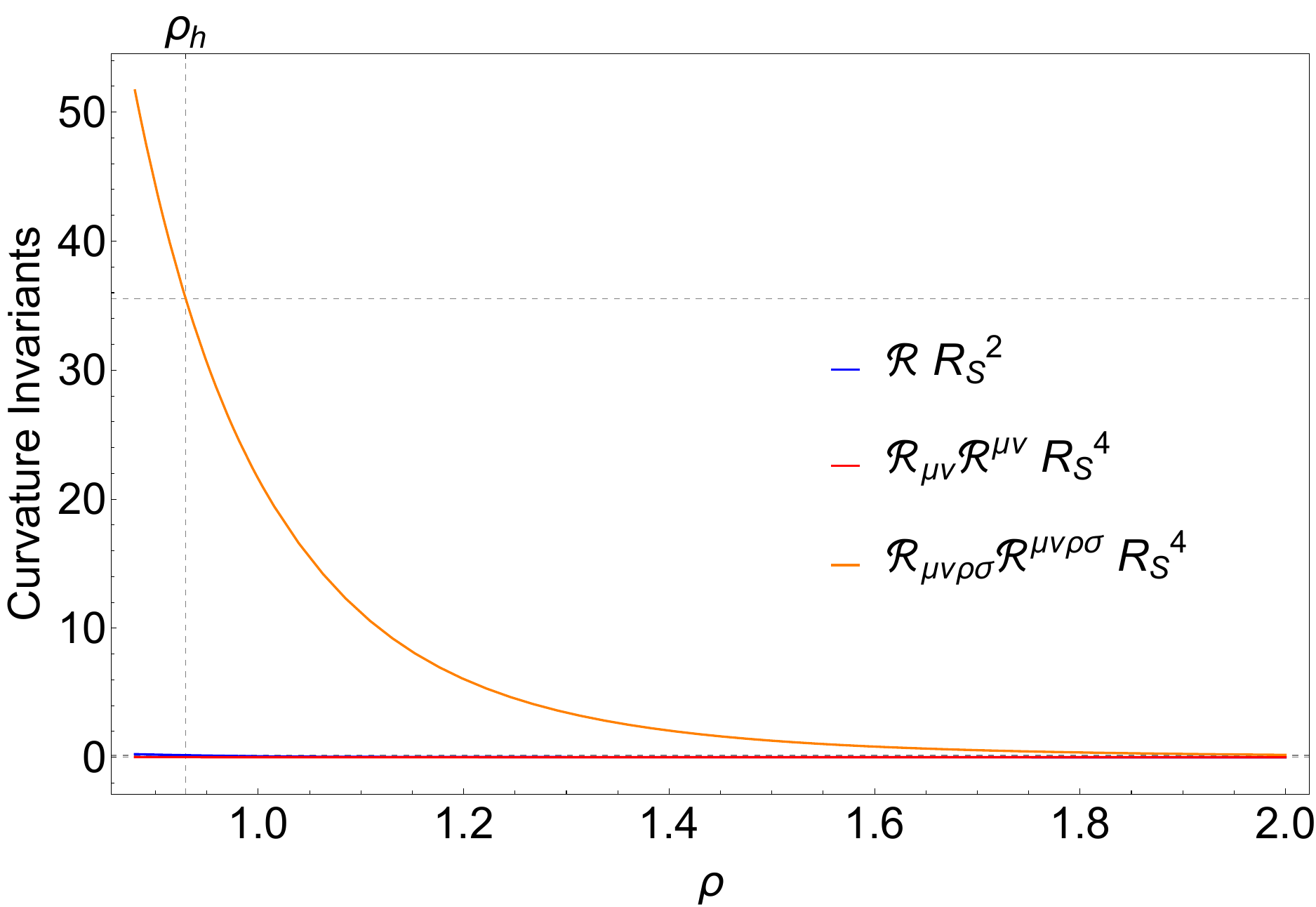}
    \caption{Plot of dimensionless curvature invariants as a function of the dimensionless coordinate $\rho$ for selected values of the parameters $\sigma = 1$ and $\mathcal{G} = 0.01$ describing a  black hole with two horizons. The invariants are plotted starting from a point at finite distance from the coordinate $\rho_{h} = 0.93$ which corresponds the outer event horizon of the black hole model. The horizontal dashed lines indicate the values of the dimensionless curvature invariants at the outer event horizon, with the position of $\rho_{h}$ marked by the vertical dashed line.}
    \label{fig:CurvatureInvariantsExternalHorizon}
\end{figure}

\begin{figure}
    \centering
    \includegraphics[width=\linewidth]{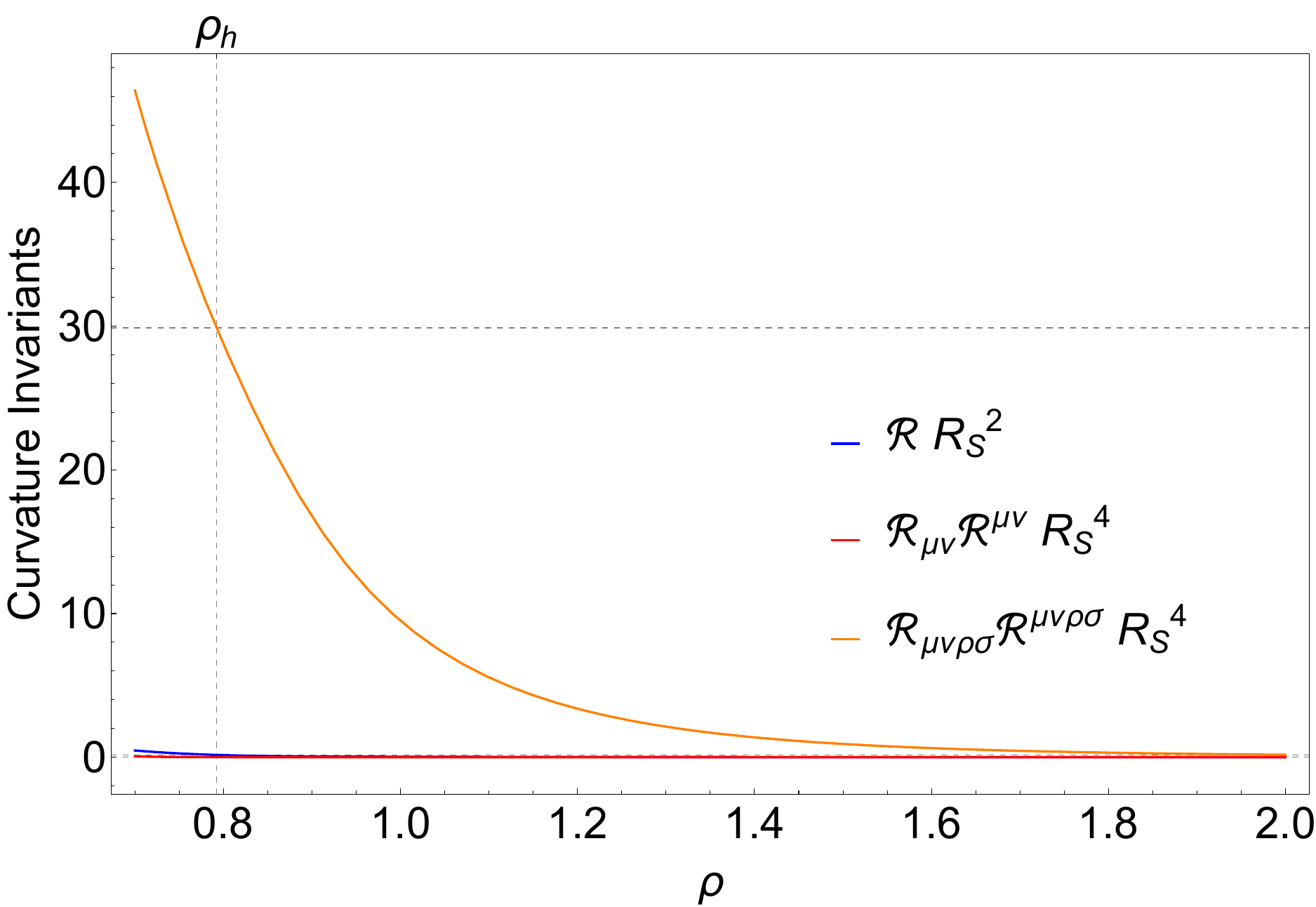}
    \caption{Plot of dimensionless curvature invariants as a function of the dimensionless coordinate $\rho$ for selected values of the parameters $\sigma = 1$ and $\mathcal{G} = 0.09$ describing an extremal black hole. The invariants are plotted from a point at finite distance from the coordinate $\rho_{h} = 0.79$, which corresponds to the degenerate event horizon of the black hole model. The horizontal dashed lines indicate the values of the dimensionless curvature invariants at the outer event horizon, with the position of $\rho_{h}$ marked by the vertical dashed line.}
    \label{fig:CurvatureInvariantsDegenerateHorizon}
\end{figure}

Therefore, the simplest way to investigate the regularity of our spacetime  at the horizons  is to check the finiteness of the dimensionless curvature invariants, $\mathcal{R} = g^{\mu\nu}R_{\mu\nu} R_\text{s}^{2}$, $\mathscr{R} = R_{\mu\nu} R^{\mu\nu} R_\text{s}^{4}$ and $\mathcal{K} = R_{\mu\nu\rho\sigma} R^{\mu\nu\rho\sigma} R_\text{s}^{4}$ at a horizon. We have computed  numerically the dimensionless curvature invariants for our numerical solutions in the parameter region for which  our solution have been obtained  using the shooting  method. We have found that  all dimensionless curvature invariants remain finite  for $\rho>0$, in particular at the outer and inner event horizons.
In \cref{fig:CurvatureInvariantsExternalHorizon,fig:CurvatureInvariantsDegenerateHorizon} we present the plots of the dimensionless curvature invariants, as functions of the dimensionless radial coordinate $\rho$, for the two selected values of the parameters used in  \cref{subsec:BHform}: a two-horizon and a one-horizon, extremal, configurations. In particular \Cref{fig:CurvatureInvariantsExternalHorizon} shows  the behavior of the dimensionless curvature invariants of the two-horizon black-hole model with $\sigma = 1$ and $\mathcal{G} = 0.01$. The $\rho$-axis has been cut at some finite distance in the black-hole interior, but we have checked that the dimensionless curvature invariants are also finite at the inner horizon. In \cref{fig:CurvatureInvariantsDegenerateHorizon}, instead, we present the same analysis for the extremal black-hole model with $\sigma = 1$ and $\mathcal{G} = 0.09$. As one can see in both plots, all dimensionless curvature invariants take finite values at the event horizon.

\section{Effective UV spacetime and linearized Renormalization-Group flow}
\label{sect:euvslrgf}
The results obtained in the previous sections were based on the analytic approximation of the RG trajectories given by \cref{RGFlowEquationApprox}.  
Although we generally expect our findings to be qualitatively independent of this approximation, it is important to verify their validity using an alternative approximation. This is especially important near the UV fixed point, where the UV geometry is quite sensitive to the shape of the coupling functions $\lambda(k)$, $g(k)$. In this section, we will repeat the calculations from \cref{UVApproximateSolutionsSection} leading to the UV quantum geometry, using a solution of the linearized RG equations near the UV fixed point rather than the analytical approximation \eqref{RGFlowEquationApprox}.

In proximity of a fixed point $u^{\alpha}_{*}$, the RG flow for the coupling constants $u^{\alpha}$ and beta functions $\beta^{\alpha}$ can be described by the linearized RG equations 
\begin{equation}
    k \partial_{k} u^{\alpha} = \sum_{\gamma} B^{\alpha}\,_{\gamma} (u^{\gamma}_{k}-u^{\gamma}_{*}) \, ,
\end{equation}
where the components of the stability matrix $B$ are given by $B^{\alpha}\,_{\gamma} \equiv \partial\beta^{\alpha}/\partial {u^{\gamma}} |_{u = u_{*}}$. 

The right-eigenvectors of $B$ are $e^{\alpha}_{i} = (\bold{e}_{i})^{\alpha}$. The solution of the linearized equation reads
\begin{equation}{\label{LinearizedSolution}}
    u^{\alpha} (k) = u_{*}^{\alpha} + \sum_{i} c_{i} e^{\alpha}_{i} \left(\frac{k}{k_{0}}\right)^{-\theta_{i}} \, ,
\end{equation}
where the $c_{i}$'s are free integration constants, $k_{0}$ is an arbitrary renormalization scale, while $\theta_{i}$ and $e^{\gamma}_{i}$ are the eigenvalues and the eigenvectors of the stability matrix $B$, respectively.  

Being the UV fixed point an attractor for the dynamical system, the coefficients $c_{i}$'s corresponding to eigenvalues $\text{Re}\, \theta_{i}<0$ must be set to zero. Note that the linearized solution \eqref{LinearizedSolution} provides a good approximation only near the fixed point, i.e., at energies $k^{2}\gtrsim m_{p}^{2}$.

When considering the Einstein-Hilbert truncation, we deal with only two coupling constants, $\lambda(k)$ and $g(k)$. It is useful to distinguish between the case where the stability coefficients $\theta_{1}$, $\theta_{2}$ are real and the case where they are complex, with $\theta_{1,2} = \theta^{\prime} \pm i \theta^{\prime\prime}$. In the former case, \cref{LinearizedSolution} takes the form
\begin{subequations}{\label{RunningEquationsRealCoeff}}
\begin{align}
    \lambda(k) & = \lambda_{*} + c_{1} e^{1}_{1} \left(\frac{k}{k_{0}}\right)^{-\theta_{1}} + c_{2} e^{1}_{2} \left(\frac{k}{k_{0}}\right)^{-\theta_{2}} \, ; \\
    g(k) & = \lambda_{*} + c_{1} e^{2}_{1} \left(\frac{k}{k_{0}}\right)^{-\theta_{1}} + c_{2} e^{2}_{2} \left(\frac{k}{k_{0}}\right)^{-\theta_{2}} \, .
\end{align}
\end{subequations}
$c_{1}$ and $c_{2}$ are the only free parameters, which select the actual RG trajectory corresponding to physical couplings. \\
In the second case -- complex eigenvalues of the stability matrix -- the eigenvectors $\bold{e}_{1,2}$ are complex conjugates and \cref{LinearizedSolution} can be written as~\cite{Bonanno:2018gck}
\begin{widetext}
    \begin{subequations}\label{RunningEquationsComplexCoeff}
    \begin{align}
            \lambda(k) & = \lambda_{*} + \left\{ \left[c_{1} \cos(\theta^{\prime\prime} t) + c_{2}\sin(\theta^{\prime\prime} t)\right] e^{1}_{1} + \left[c_{1} \cos(\theta^{\prime\prime} t) - c_{2}\sin(\theta^{\prime\prime} t)\right] e^{1}_{2}\right\} \left(\frac{k}{k_{0}}\right)^{-\theta^{\prime}} \, ,\\
            g(k) & = g_{*} + \left\{ \left[c_{1} \cos(\theta^{\prime\prime} t) + c_{2}\sin(\theta^{\prime\prime} t)\right] e^{2}_{1} + \left[c_{1} \cos(\theta^{\prime\prime} t) - c_{2}\sin(\theta^{\prime\prime} t)\right] e^{2}_{2}\right\} \left(\frac{k}{k_{0}}\right)^{-\theta^{\prime}} \, ,
    \end{align}
\end{subequations}
\end{widetext}
where $t = \ln(k/k_{0})$. $k_{0}$ can be identified as the Newton constant $G_{0} = \ell_{p}^{2}$. 

The actual values of the relevant parameters $\theta_{i}$, $e^{2}_{1}$, $e^{1}_{2}$ \footnote{The following normalization is a common choice: $e^{1}_{1} = e^{2}_{2} = 1$.} depend on the number of fields in the non-gravitational sector and on the gauge choice for the gravitational sector (see Refs.~\cite{Bonanno:2018gck,Biemans:2017zca,Alkofer:2018fxj} for details). 

Using the same procedure of \cref{UVApproximateSolutionsSection}, we can now derive the effective UV quantum spacetimes for the gravity-matter systems discussed in~\cite{Bonanno:2018gck} using the linearised approximations. This is achieved by using either \cref{RunningEquationsRealCoeff} or \cref{RunningEquationsComplexCoeff} instead of \cref{RGFlowEquationApprox}, with the values of $\theta_{i}$, $e^{2}_{1}$, $e^{1}_{2}$ given by~\cite{Bonanno:2018gck}.

The primarly difference between the approximate solutions of the RG equations provided 
by \cref{RGFlowEquationApprox,RunningEquationsRealCoeff,RunningEquationsComplexCoeff} is as follows. 
\Cref{RGFlowEquationApprox} provides an analytical approximation of the full RG flow, preserving the information about the (non-linear) coupling between $\lambda(k)$ and $g(k)$. Conversely, \cref{RunningEquationsRealCoeff,RunningEquationsComplexCoeff} give an approximate linear solution of the RG flow near the UV NGFP. In this approximation, the flow of $\lambda(k)$ and $g(k)$ is decoupled and the information about the RG trajectory is entirely captured by the eigenvalues and eigenvectors $\theta_{i}$, $e^{2}_{1}$, $e^{1}_{2}$ of the stability matrix. We will see below that these differences have also important consequences on the shape of the UV quantum geometry. 

The UV-approximate solutions corresponding to the linearised RG flow can now be found using the same iterative procedure performed in \cref{UVApproximateSolutionsSection}. We will first consider the case of  real eigenvalues $\theta_{i}$'s and then briefly comment the case of complex eigenvalues. Notice that covariant pure gravity corresponds to real values of the $\theta_i$'s for  all the gravity-matter systems considered in Ref.~\cite{Bonanno:2018gck}.
Complex $\theta_{i}$'s values occur only in the case of foliated pure gravity.
Inserting \cref{RunningEquationsRealCoeff} into \cref{MetricFunctionDressed} we get the dressed metric function
\begin{widetext}
    \begin{equation}
    \begin{split}\label{eqn2}
            f_{\text{d}}(r)  =& 1 -\frac{2 M \left\{g_{*} + c_{1} e^{2}_{1} \left[G_{0} k(r)\right]^{-\theta_{1}}+c_{2} e^{2}_{2} \left[G_{0} k(r)\right]^{-\theta_{2}}\right\}}{r k(r)^2}  \\
            & -\frac{1}{3} r^2 k(r)^2 \left\{\lambda_{*} + c_{1} e^{1}_{1} \left[G_{0} k(r)\right]^{-\theta_{1}}+c_{2} e^{1}_{2} \left[G_{0} k(r)\right]^{-\theta_{2}}\right\} \, .
    \end{split}
\end{equation}
\end{widetext}
Following the approach in \cref{UVApproximateSolutionsSection}, we now expand $k(r)$ as a power series near $r=0$, $k(r)^{2} = \sum_{n=-N}^{\infty} a_{n} r^{n}$, and solve \cref{GeodesicEquation1} using the Frobenius method. For $N$ we find here the same value found in 
\cref{UVApproximateSolutionsSection}, i.e., $N=2$, so that the expansion for $k(r)$ has the same form given in \cref{kUV}.
\Cref{GeodesicEquation1} includes not only terms of order $r^m,\, m=0,1,2$ terms, but also contributions of order $r^{1+\theta_1},\,r^{2+\theta_1},r^{1+\theta_2},\,r^{2+\theta_2} $ and higher. 
However, for all the matter-gravity systems considered with real $\theta_i$'s, we have $\theta_i>2$ (see Ref.~\cite{Bonanno:2018gck}) making these terms subleading compared to the $r^m$ terms with $m=0,1,2$. As a result, the coefficients $\xi$, $\beta$, $\gamma$, $\delta$ in the expansion \eqref{kUV} are fully determined by the $r^m,\, m=0,1,2$ terms in the Frobenius expansion of the non-linear ODE \eqref{GeodesicEquation1}. We get

\begin{subequations}{\label{eqn1}}
    \begin{align}
        \xi & = \frac{1}{\sqrt{\lambda_{*}}}\frac{\sqrt{3} \sqrt{\sigma }}{\sqrt{3-\sigma}} \, ;\\
        \gamma & = \frac{6 g_{*} M}{\sigma - 6} \, ;\\
        \delta & = - \frac{9 g_{*}^{2} \lambda_{*} M^{2}}{\sigma (6 - \sigma)^{2}} \, , \label{eqn1_delta}
    \end{align}
\end{subequations}
where, as usual, we have defined $\beta \lambda_{*} = \sigma$ with $0<\sigma<3$.\\
The parameters $\xi$ and $\gamma$ retain the same values found in \cref{UVApproximateSolutionsSection}, but $\delta$ takes a different value (see \cref{kUVc}). In particular, in the present case, $\delta$ depends solely on the black hole mass $M$ and is independent of the Planck mass. 

Using the values of the parameters \eqref{eqn1} into \cref {eqn2}, we get the UV form of the metric function \eqref{DressedMetricFunctionUV}
\begin{equation}{\label{LinearizedMetricFunctionUV}}
    f_{\text{d}}^{(\text{UV})} = \left( 1- \frac{\sigma}{3} \right) - \omega r - \frac{r^{2}}{L^{2}} + \mathcal{O}(r^{3}) \, ,
\end{equation}
but here
\begin{equation}
    \omega = \frac{4 \mathcal{M} (\sigma - 3)}{\sigma  (6-\sigma)}\, , \quad \frac{1}{L^{2}} = \frac{3 \mathcal{M}^{2} (5 \sigma - 24)}{\sigma ^2 (\sigma-6)^2} \, ,
\end{equation}
where $\mathcal{M} = g_{*}\lambda_{*} M$.
Comparing this expression for the dressed metric function with that given in \cref{DressedMetricFunctionUV}, one can easily realize not only that it has exactly the same form found in \cref{UVApproximateSolutionsSection}, but also that the values of the conical singularity and the linear term coefficient remain unchanged. Thus, the linearized FRG flow generates the same spacetime singularities as we have found using an analytic approximation to the coupling functions.

However, there is a crucial difference between \cref{LinearizedMetricFunctionUV} and \cref{DressedMetricFunctionUV}.
While in the latter case, the coefficient of the $r^2$ term, i.e., $1/L^2$, can change sign, producing the AdS/dS phase transition discussed above, in the former case it is always negative, resulting in a UV geometry that remains dS. 

The absence of an AdS/dS phase transition in the UV geometry \eqref{LinearizedMetricFunctionUV} arises as consequence of the linear approximation used for the RG flow near the UV fixed point. We recall that the transition is driven by the competition between the Plank mass $\mathit{m}_{p}$ and the black-hole mass $M$ (see \cref{tp1}). When using the linear approximation, however, the coefficient \eqref{eqn1_delta} no longer depend on $\mathit{m}_{p}$, and thus the latter drops out from $1/L^2$. This can also be seen by expanding $\Lambda_{k}$ for $k\to \infty$ from \cref{RGFlowEquationApprox}:
\begin{equation}
    \Lambda_{k} \sim \lambda_{*}  k(r)^2-\frac{7 \mathcal{M}_{p}^{2}}{2} + \mathcal{O}\left[k(r)^{-1}\right] \, .
\end{equation}
The last term in the previous expression gives the $\mathit{m}_{p}$-dependent contribution to $1/L^2$, which is absent in the linear approximation of the UV RG flow.

Finally, let us conclude this section with some comments regarding the case of complex eigenvalues $\theta_i$'s of the stability matrix. In this case, we need to use \cref{RunningEquationsComplexCoeff} into \cref{MetricFunctionDressed}, instead of \cref{RunningEquationsRealCoeff}. The resulting expression for the effective metric function has the same form found for the real $\theta_i$ case, i.e., a sum of powers $r^m,\, m=0,1,2$ and $r^{1+\theta'},\,r^{2+\theta'},r^{1+\theta''},\,r^{2+\theta''} $.  However, in the classification of Ref.~\cite{Bonanno:2018gck}, the case of complex eigenvalues describes foliated pure gravity and corresponds to (non-rational) values of $\theta'\sim 0.2$. This gives non-analytic, leading terms like $r^{1+\theta'}$, which seems a distinguishing  feature of  foliated pure gravity. 

\section{Conformal sector and anomalous dimension}
\label{sect:csad}
In the context of asymptotic safe gravity, an important aspect is encoded in the scaling dimension of the Newton constant $G_N$. The inverse of $G_{N}$ is equivalent to the wave-function renormalization function $Z_{N}$ of the graviton, which should scale as the inverse of the cutoff squared~\cite{Percacci:2004sb}.
Accordingly, the anomalous dimension of the field is given by the relation $ \tilde\eta_N = \beta_N -2$, where $\beta_N$ denotes the beta-function of the Newton constant. At a fixed point where $\beta_N$ vanishes, the anomalous dimension $\tilde\eta_N$ takes the value $\tilde\eta_N = -2$, reflecting the classical scaling dimension of the background Newton constant.\\

It has been argued that the dynamical information about the anomalous dimension is better described by the anomalous dimension $\eta_{N}$ associated with graviton fluctuations~\cite{Percacci:2004sb, Bonanno:2021squ, Bonanno:2023fij}, which is expected to take values closer to $1$ ($1.02$ according to Ref.~\cite{Bonanno:2021squ}.) The determination of $\eta_{N}$ does not rely on the $\beta$-function approach, which inherently involves some truncation scheme in the field theory space. Following Refs.~\cite{Bonanno:2021squ,Bonanno:2023fij}, starting from the action \eqref{EHT}, the dynamical fluctuations of the gravitational field can be described in terms of the conformal mode of the metric $\phi$. This is defined through a Weyl rescaling from $g_{\mu\nu}$ to a reference metric $\hat g_{\mu\nu}$: $g_{\mu\nu}=\phi^2\hat g_{\mu\nu}$. From \cref{gs},  an appropriate choice of the reference metric $\hat g_{\mu\nu}$, allows the identification of  the conformal mode in terms of the metric function, given by $\phi= \sqrt{f}$. Using the proper distance formula \eqref{GeodesicEquation},  the anomalous dimension of the conformal factor of the metric can be determined from the dimensionless proper length $\mathcal{D}$ in equation \cref{rv} and reads
\begin{equation}
    \eta_{N} = 1 - \left[\mathcal{D}\right]\, ,
\end{equation}
where $\left[\mathcal{D}\right]$ is the scaling dimension of $\mathcal{D}$, given by
\begin{equation}
    \left[\mathcal{D}\right] = \frac{\rho}{\mathcal{D}} \frac{\dd \mathcal{D}}{\dd \rho}\, .
\end{equation}
The anomalous dimension $\eta_{N}$ will be therefore a function of the radial coordinate $r$, reflecting its running with the momentum scale $k$.  
Using the numerical results obtained from our shooting procedure, we evaluate $\eta_{N}(\rho)$ for several values of the parameters $\sigma$ and $\mathcal{G}$. As expected, while the qualitative behavior of $\eta_{N}(\rho)$ remains largely unaffected by changes in $\sigma$, it exhibits a significant dependence on the mass $M$.  
  
\begin{figure}
    \centering
    \includegraphics[width=\linewidth]{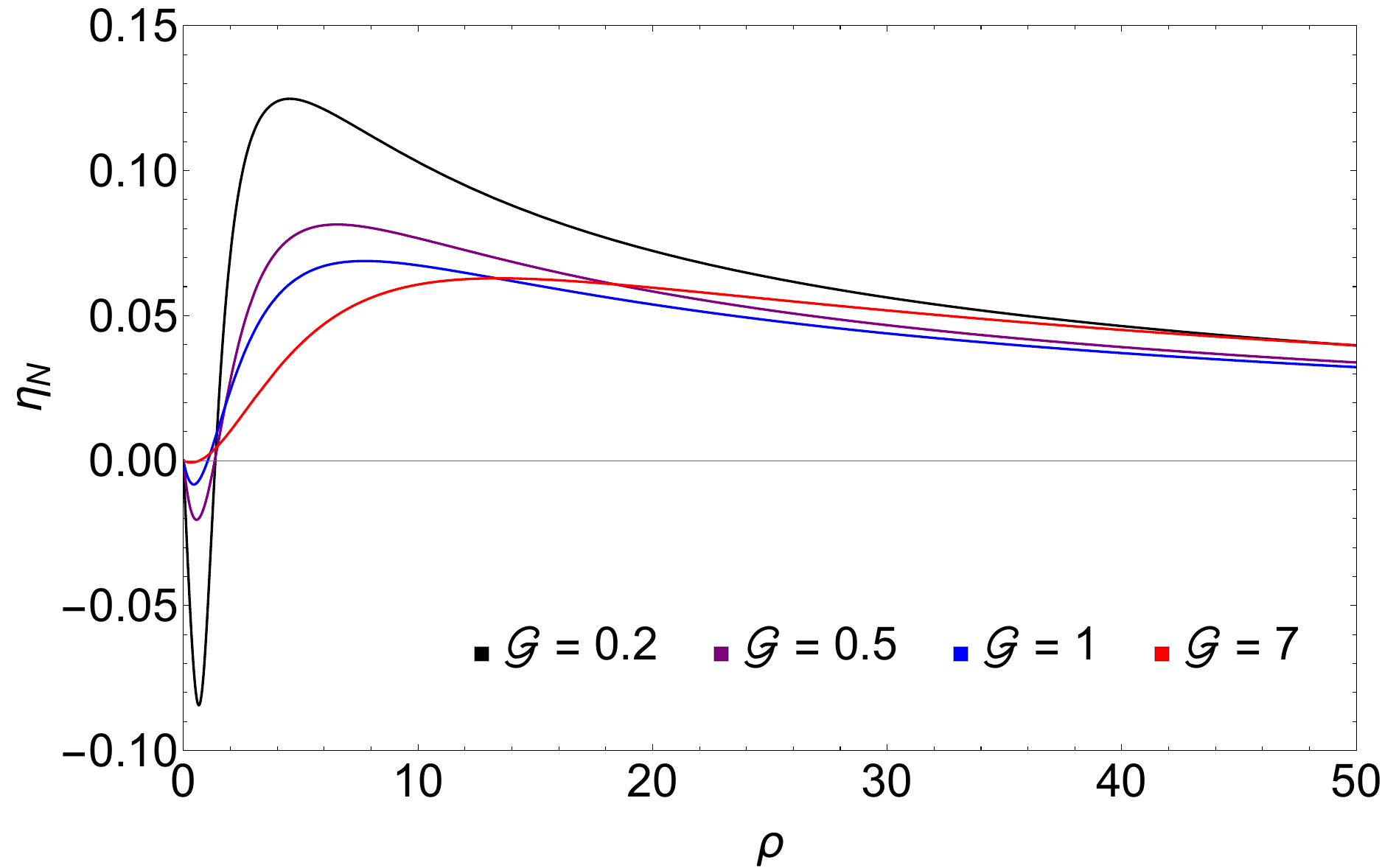}
    \caption{Plots of the anomalous dimension  $\eta_{N}(\rho) $ over the interval $\rho\in(0,\,50]$ with a fixed value $\sigma = 1$ and varying $\mathcal{G}$ from $\mathcal{G}=0.2$ to $\mathcal{G}=7$.}
    \label{fig:Detail_Anomalous_Dimension_Shooting}
\end{figure}

Figure \cref{fig:Detail_Anomalous_Dimension_Shooting} illustrates the typical behavior of $\eta_{N}(\rho)$. We have always $|\eta_{N}|<1$, and it can take both positive and negative values. Notably, $\eta_{N}(\rho)$ approaches zero in both the far IR (small $\rho$) and in the deep UV (large  $\rho$), near the NGFP. This is a consequence of the linear behavior $\mathscr{L}\sim r$ near $r=0$ and $r\to \infty$. While $\eta_{N}(\rho)$ decreases monotonically towards zero at large $\rho$, its behavior at small $\rho$ is quite non-trivial.

\begin{figure}
    \centering
    \includegraphics[width=\linewidth]{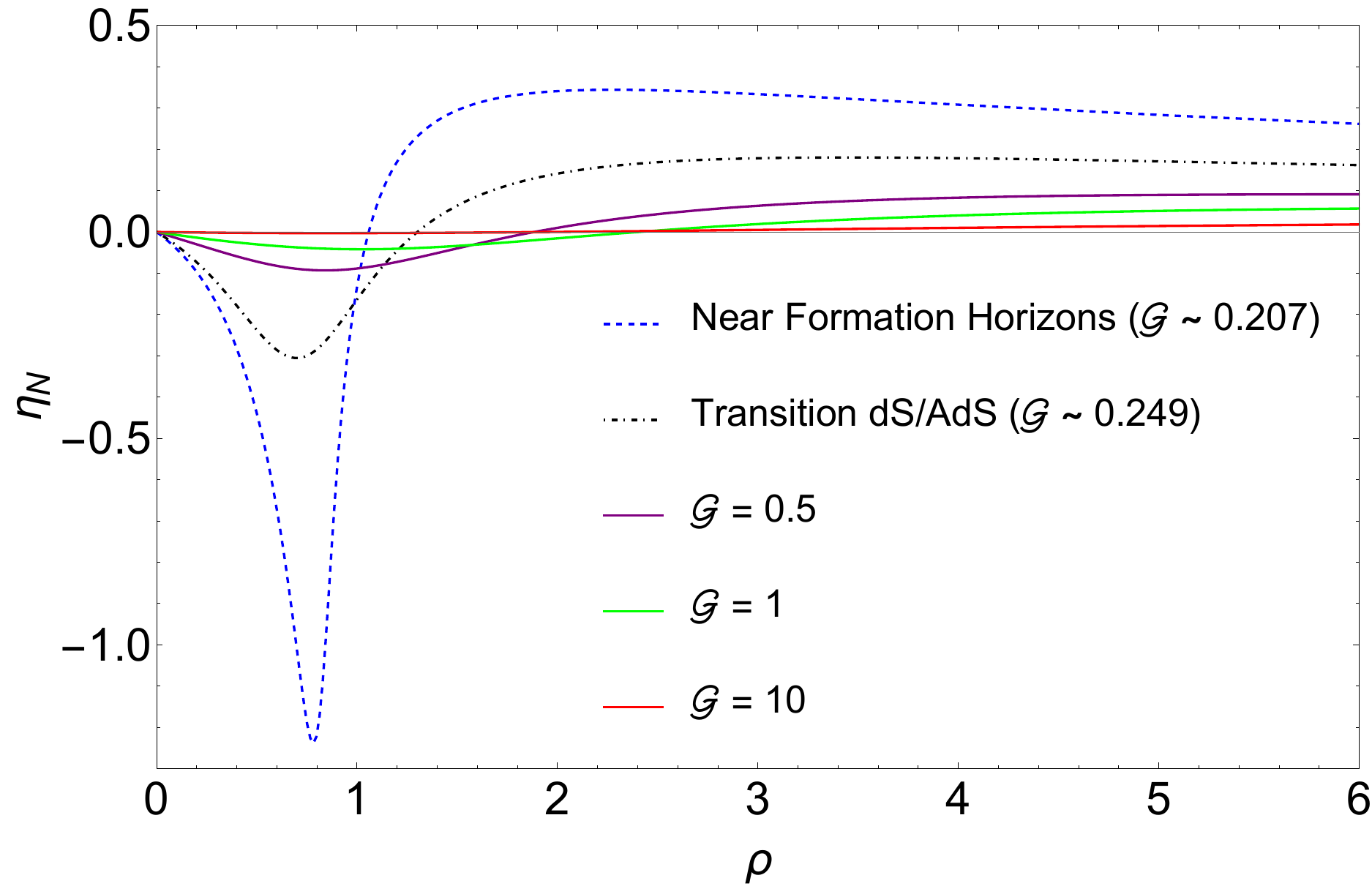}
    \caption{Plots of the anomalous dimension $\eta_{N}(\rho)$ over the interval $\rho\in(0,\,6]$, for $\sigma = 0.5$. We compare the behavior of $\eta$ for $\mathcal{G}\sim 0.207$ (dashed blue curve), when horizons are about to form and for $\mathcal{G}\sim 0.249$ and at the AdS/dS transition. }
    \label{fig:Anomalous_Dimension_G_0._5}
\end{figure}

Figures \cref{fig:Detail_Anomalous_Dimension_Shooting,fig:Anomalous_Dimension_G_0._5} show plots    of $\eta_{N}(\rho)$ within the range $\rho\in(0,\,50]$  and $\rho\in(0,\,6]$, respectively. Starting from $\eta_{N} = 0$ at the NGFP  the anomalous dimension initially decreases, reaching negative values for small $\rho$. It attains a minimum that can be quite low for large values of the mass $M$. The curve then rises, crosses zero around $\rho\sim 1$ (corresponding to physical distances on the order of the Schwarzschild radius), until it reaches a positive maximum.  

Finally, in figure \cref{fig:Anomalous_Dimension_G_0._5} we illustrate the impact of the $\text{AdS}$/$\text{dS}$ and black hole transitions on the $\eta_{N}(\rho)$
curve.
For small sub-Planckian values of the mass $M$ (corresponding to large  $\mathcal{G}$) the anomalous dimension remains quite close to zero. However, as $M$ increases and approaches the $\text{AdS}$/$\text{dS}$ transition, deviations from zero becomes significant. These deviations grow even more relevant as the mass is further increased, nearing the black hole transition.

The results presented here for the anomalous dimension $\eta_{N}$ are in agreement with those of~\cite{Bonanno:2023fij}. In that paper it was found that, for  $\eta_{N}<\eta_c\approx 0.96$ there exists a line of FPs with the desired physical properties (positive Newton's and cosmological constant, presence of complex eigenvalues with positive real part producing the spiral behavior near the UV fixed point). Conversely, for $\eta_{N}>1$ the theory exhibit  an infinite number of relevant directions making it  non-predictive .   

\section{Conclusions}
\label{sect:c}

In this paper, we have constructed  effective quantum spacetime geometries by deforming the classical SdS solution using an FRG approach. Unlike previous efforts along  this direction, our approach is fully  self-consistent, as  it takes in full account the  feed-back of the geometry in determining the dependence of the cutoff scale $k$ on the radial coordinate $r$. This self-consistency enabled us to derive the analytic forms for the deformed metric solution in both the UV and in the IR regimes as well as to numerically determine a global interpolating solution.  

The most significant new results emerge from the  short-distance behavior of the FRG-deformed quantum geometry. The UV phase structure we uncovered is remarkably  rich and supports recent studies exploring the UV critical manifolds within the FRG framework. Our findings reveal that the quantum corrections remove the classical Schwarzschild singularity  replacing it with  a conical singularity.
Moreover, we identified two critical values of the mass, $M_{t}$ and $M_{c}$, both of order of the Planck mass, corresponding, respectively, to a transition AdS/dS spacetime and to the formation of black hole horizons. The non-trivial nature of the UV spacetime is further corroborated by the behavior of the anomalous dimension in the UV. Finally, we verified that the black-hole spacetimes remain regular at the event horizon. However, the metric function near the event horizons  seems to have a non-trivial, non analytic behaviour, whose origin we argue arises from the interpolation between the UV and IR physics inherent to the construction of the quantum-effective solution. This  non-trivial non-analytic behaviour  of the metric function at the black hole's event horizon can thus be traced back directly to the quantum deformations induced by the running of the coupling constants $G_{k}$ and $\Lambda_{k}$.  These quantum-induced effects might have phenomenological implications for astrophysical black hole, potentially leading to observable signatures in gravitational waves and quasinormal mode spectra.

The overall picture that emerges can be summarized as follows. \\ 
$\bold{(a)} \,M<M_{t}.\, $  In this regime, at short distances, we have a conical singularity, the matter contributions are subleading and the cosmological constant is dominated by the negative energy of the vacuum. The conical singularity indicates the presence of a sub-Planckian, pre-geometric phase of unbroken diffeomorphism invariance characterized by a vanishing vacuum expectation value of the conformal factor of the metric, $\langle \phi \rangle = 0$. This pre-geometric phase is also characterized by conformal unbroken symmetry, as strongly suggested by the presence of the NGFP and by the agreement of our UV metric solution with that of conformal Weyl gravity.
$\langle \phi \rangle \neq 0$, instead, generates the geometric phase in which the metric is given by \cref{DressedMetricFunctionUV}, and both diffeomorphism and conformal invariance are broken. In this phase, the spacetime develops a conical singularity,  behaves at short distances  as $\text{AdS}_{2}\times \text{S}^2$ and becomes Schwarzschild at large distances.\\
$\bold{(b)} \,M_{t}<M<M_{c}.\, $When the mass $M$ exceeds the first critical value $M_{t}$, the cosmological constant becomes dominated by matter, which gives a positive energy contributions. The spacetime interpolates between a $\text{dS}_{2}\times \text{S}^2$ throat at short distances to an asymptotically-flat spacetime at $r\to \infty$. As in the previous case, the throat terminates in a conical singularity at $r=0$. However, the low value of the mass does not allow for the formation of horizons.\\
$\bold{(c)} \,M>M_{c}.\,$ When the mass reaches the second critical value $M_{c}$, the short distance geometry is still described as a $\text{dS}_2$ $\times \text{S}^2$ throat, ending in a conical singularity. However, at this stage, the mass is large  enough to allow  for the formation of horizons.  Globally, this configuration closely resembles  black holes with a dS core~\cite{Cadoni:2023nrm}, differing primarily in the $r\sim 0$ region.

The transition between regimes $(b)$ and $(c)$ appears to be a Planck-scale analog  of the formation mechanism of stellar black holes. It is characterized by the competition between the attractive force, generated by matter, and the repulsive quantum-gravity effects,  generating a (Planck-scale) threshold mass for  black-hole formation.

Regarding  the IR behavior of our FRG quantum-improved black hole solutions, we found only slight deviations from the findings in Ref.~\cite{Bonanno:2000ep}.
Asymptotically, as $r\to \infty$, the solution behaves, at leading order, as classical Schwarzschild solution, with subleading contribution accounting for quantum Planckian hair.

The most notable new features are presence of new ($1/r^2$  and $\ln r/r^3$) hair generated by the IR flow of the cosmological constant and   the potential presence of  a super-Planckian $1/r^3$ hair. 
In this paper we did not delve into this matter as we simply assumed, inspired by a naturalness criteria, that this hair is also of Planckian scale. However, new physics at horizon scales and/or inclusion of non-local effects in the IR regime of the FRG scenario could be responsible for the generation of such a super-Planckian hair. This latter possibility is quite appealing from a phenomenological point of view, as we expect super-Planckian hairs could produce  observational signatures detectable by current  astrophysical observational facilities~\cite{Cadoni:2022vsn}.

\section*{Acknowledgements}
We are grateful to Mauro Oi for useful discussions on the shooting method.

\newpage

\bibliography{References}

\begin{thebibliography}{35}%
\makeatletter
\providecommand \@ifxundefined [1]{%
 \@ifx{#1\undefined}
}%
\providecommand \@ifnum [1]{%
 \ifnum #1\expandafter \@firstoftwo
 \else \expandafter \@secondoftwo
 \fi
}%
\providecommand \@ifx [1]{%
 \ifx #1\expandafter \@firstoftwo
 \else \expandafter \@secondoftwo
 \fi
}%
\providecommand \natexlab [1]{#1}%
\providecommand \enquote  [1]{``#1''}%
\providecommand \bibnamefont  [1]{#1}%
\providecommand \bibfnamefont [1]{#1}%
\providecommand \citenamefont [1]{#1}%
\providecommand \href@noop [0]{\@secondoftwo}%
\providecommand \href [0]{\begingroup \@sanitize@url \@href}%
\providecommand \@href[1]{\@@startlink{#1}\@@href}%
\providecommand \@@href[1]{\endgroup#1\@@endlink}%
\providecommand \@sanitize@url [0]{\catcode `\\12\catcode `\$12\catcode `\&12\catcode `\#12\catcode `\^12\catcode `\_12\catcode `\%12\relax}%
\providecommand \@@startlink[1]{}%
\providecommand \@@endlink[0]{}%
\providecommand \url  [0]{\begingroup\@sanitize@url \@url }%
\providecommand \@url [1]{\endgroup\@href {#1}{\urlprefix }}%
\providecommand \urlprefix  [0]{URL }%
\providecommand \Eprint [0]{\href }%
\providecommand \doibase [0]{http://dx.doi.org/}%
\providecommand \selectlanguage [0]{\@gobble}%
\providecommand \bibinfo  [0]{\@secondoftwo}%
\providecommand \bibfield  [0]{\@secondoftwo}%
\providecommand \translation [1]{[#1]}%
\providecommand \BibitemOpen [0]{}%
\providecommand \bibitemStop [0]{}%
\providecommand \bibitemNoStop [0]{.\EOS\space}%
\providecommand \EOS [0]{\spacefactor3000\relax}%
\providecommand \BibitemShut  [1]{\csname bibitem#1\endcsname}%
\let\auto@bib@innerbib\@empty
\bibitem [{\citenamefont {Dittrich}\ and\ \citenamefont {Reuter}(1985)}]{Dittrich:1985yb}%
  \BibitemOpen
  \bibfield  {author} {\bibinfo {author} {\bibfnamefont {W.}~\bibnamefont {Dittrich}}\ and\ \bibinfo {author} {\bibfnamefont {M.}~\bibnamefont {Reuter}},\ }\href@noop {} {\emph {\bibinfo {title} {{EFFECTIVE LAGRANGIANS IN QUANTUM ELECTRODYNAMICS}}}},\ Vol.\ \bibinfo {volume} {220}\ (\bibinfo {year} {1985})\BibitemShut {NoStop}%
\bibitem [{\citenamefont {Nielsen}(1981)}]{Nielsen:1980sx}%
  \BibitemOpen
  \bibfield  {author} {\bibinfo {author} {\bibfnamefont {N.~K.}\ \bibnamefont {Nielsen}},\ }\href {\doibase 10.1119/1.12565} {\bibfield  {journal} {\bibinfo  {journal} {Am. J. Phys.}\ }\textbf {\bibinfo {volume} {49}},\ \bibinfo {pages} {1171} (\bibinfo {year} {1981})}\BibitemShut {NoStop}%
\bibitem [{\citenamefont {Polyakov}(1993)}]{Polyakov:1993tp}%
  \BibitemOpen
  \bibfield  {author} {\bibinfo {author} {\bibfnamefont {A.~M.}\ \bibnamefont {Polyakov}},\ }in\ \href@noop {} {\emph {\bibinfo {booktitle} {{Les Houches Summer School on Gravitation and Quantizations, Session 57}}}}\ (\bibinfo {year} {1993})\ pp.\ \bibinfo {pages} {0783--804},\ \Eprint {http://arxiv.org/abs/hep-th/9304146} {arXiv:hep-th/9304146} \BibitemShut {NoStop}%
\bibitem [{\citenamefont {Reuter}(1998)}]{Reuter:1996cp}%
  \BibitemOpen
  \bibfield  {author} {\bibinfo {author} {\bibfnamefont {M.}~\bibnamefont {Reuter}},\ }\href {\doibase 10.1103/PhysRevD.57.971} {\bibfield  {journal} {\bibinfo  {journal} {Phys. Rev. D}\ }\textbf {\bibinfo {volume} {57}},\ \bibinfo {pages} {971} (\bibinfo {year} {1998})},\ \Eprint {http://arxiv.org/abs/hep-th/9605030} {arXiv:hep-th/9605030} \BibitemShut {NoStop}%
\bibitem [{\citenamefont {Codello}\ \emph {et~al.}(2008)\citenamefont {Codello}, \citenamefont {Percacci},\ and\ \citenamefont {Rahmede}}]{Codello:2007bd}%
  \BibitemOpen
  \bibfield  {author} {\bibinfo {author} {\bibfnamefont {A.}~\bibnamefont {Codello}}, \bibinfo {author} {\bibfnamefont {R.}~\bibnamefont {Percacci}}, \ and\ \bibinfo {author} {\bibfnamefont {C.}~\bibnamefont {Rahmede}},\ }\href {\doibase 10.1142/S0217751X08038135} {\bibfield  {journal} {\bibinfo  {journal} {Int. J. Mod. Phys. A}\ }\textbf {\bibinfo {volume} {23}},\ \bibinfo {pages} {143} (\bibinfo {year} {2008})},\ \Eprint {http://arxiv.org/abs/0705.1769} {arXiv:0705.1769 [hep-th]} \BibitemShut {NoStop}%
\bibitem [{\citenamefont {Falls}\ \emph {et~al.}(2016)\citenamefont {Falls}, \citenamefont {Litim}, \citenamefont {Nikolakopoulos},\ and\ \citenamefont {Rahmede}}]{Falls:2014tra}%
  \BibitemOpen
  \bibfield  {author} {\bibinfo {author} {\bibfnamefont {K.}~\bibnamefont {Falls}}, \bibinfo {author} {\bibfnamefont {D.~F.}\ \bibnamefont {Litim}}, \bibinfo {author} {\bibfnamefont {K.}~\bibnamefont {Nikolakopoulos}}, \ and\ \bibinfo {author} {\bibfnamefont {C.}~\bibnamefont {Rahmede}},\ }\href {\doibase 10.1103/PhysRevD.93.104022} {\bibfield  {journal} {\bibinfo  {journal} {Phys. Rev. D}\ }\textbf {\bibinfo {volume} {93}},\ \bibinfo {pages} {104022} (\bibinfo {year} {2016})},\ \Eprint {http://arxiv.org/abs/1410.4815} {arXiv:1410.4815 [hep-th]} \BibitemShut {NoStop}%
\bibitem [{\citenamefont {Nink}\ and\ \citenamefont {Reuter}(2013)}]{Nink:2012vd}%
  \BibitemOpen
  \bibfield  {author} {\bibinfo {author} {\bibfnamefont {A.}~\bibnamefont {Nink}}\ and\ \bibinfo {author} {\bibfnamefont {M.}~\bibnamefont {Reuter}},\ }\href {\doibase 10.1007/JHEP01(2013)062} {\bibfield  {journal} {\bibinfo  {journal} {JHEP}\ }\textbf {\bibinfo {volume} {01}},\ \bibinfo {pages} {062} (\bibinfo {year} {2013})},\ \Eprint {http://arxiv.org/abs/1208.0031} {arXiv:1208.0031 [hep-th]} \BibitemShut {NoStop}%
\bibitem [{\citenamefont {Bonanno}\ and\ \citenamefont {Reuter}()}]{Bonanno:2000ep}%
  \BibitemOpen
  \bibfield  {author} {\bibinfo {author} {\bibfnamefont {A.}~\bibnamefont {Bonanno}}\ and\ \bibinfo {author} {\bibfnamefont {M.}~\bibnamefont {Reuter}},\ }\href {\doibase 10.1103/PhysRevD.62.043008} {\bibfield  {journal} {\bibinfo  {journal} {\href{http://dx.doi.org/10.1103/PhysRevD.62.043008} {Phys. Rev. D \textbf{62} (2000), 043008}}\ }10.1103/PhysRevD.62.043008},\ \Eprint {http://arxiv.org/abs/hep-th/0002196} {arXiv:hep-th/0002196} \BibitemShut {NoStop}%
\bibitem [{\citenamefont {Cadoni}\ \emph {et~al.}(2023{\natexlab{a}})\citenamefont {Cadoni}, \citenamefont {Sanna}, \citenamefont {Pitzalis}, \citenamefont {Banerjee}, \citenamefont {Murgia}, \citenamefont {Hazra},\ and\ \citenamefont {Branchesi}}]{Cadoni:2023lum}%
  \BibitemOpen
  \bibfield  {author} {\bibinfo {author} {\bibfnamefont {M.}~\bibnamefont {Cadoni}}, \bibinfo {author} {\bibfnamefont {A.~P.}\ \bibnamefont {Sanna}}, \bibinfo {author} {\bibfnamefont {M.}~\bibnamefont {Pitzalis}}, \bibinfo {author} {\bibfnamefont {B.}~\bibnamefont {Banerjee}}, \bibinfo {author} {\bibfnamefont {R.}~\bibnamefont {Murgia}}, \bibinfo {author} {\bibfnamefont {N.}~\bibnamefont {Hazra}}, \ and\ \bibinfo {author} {\bibfnamefont {M.}~\bibnamefont {Branchesi}},\ }\href {\doibase 10.1088/1475-7516/2023/11/007} {\bibfield  {journal} {\bibinfo  {journal} {JCAP}\ }\textbf {\bibinfo {volume} {11}},\ \bibinfo {pages} {007} (\bibinfo {year} {2023}{\natexlab{a}})},\ \Eprint {http://arxiv.org/abs/2306.11588} {arXiv:2306.11588 [gr-qc]} \BibitemShut {NoStop}%
\bibitem [{\citenamefont {Cadoni}\ \emph {et~al.}(2024)\citenamefont {Cadoni}, \citenamefont {Murgia}, \citenamefont {Pitzalis},\ and\ \citenamefont {Sanna}}]{Cadoni:2023lqe}%
  \BibitemOpen
  \bibfield  {author} {\bibinfo {author} {\bibfnamefont {M.}~\bibnamefont {Cadoni}}, \bibinfo {author} {\bibfnamefont {R.}~\bibnamefont {Murgia}}, \bibinfo {author} {\bibfnamefont {M.}~\bibnamefont {Pitzalis}}, \ and\ \bibinfo {author} {\bibfnamefont {A.~P.}\ \bibnamefont {Sanna}},\ }\href {\doibase 10.1088/1475-7516/2024/03/026} {\bibfield  {journal} {\bibinfo  {journal} {JCAP}\ }\textbf {\bibinfo {volume} {03}},\ \bibinfo {pages} {026} (\bibinfo {year} {2024})},\ \Eprint {http://arxiv.org/abs/2309.16444} {arXiv:2309.16444 [gr-qc]} \BibitemShut {NoStop}%
\bibitem [{\citenamefont {Pawlowski}\ and\ \citenamefont {Reichert}(2021)}]{Pawlowski:2020qer}%
  \BibitemOpen
  \bibfield  {author} {\bibinfo {author} {\bibfnamefont {J.~M.}\ \bibnamefont {Pawlowski}}\ and\ \bibinfo {author} {\bibfnamefont {M.}~\bibnamefont {Reichert}},\ }\href {\doibase 10.3389/fphy.2020.551848} {\bibfield  {journal} {\bibinfo  {journal} {Front. in Phys.}\ }\textbf {\bibinfo {volume} {8}},\ \bibinfo {pages} {551848} (\bibinfo {year} {2021})},\ \Eprint {http://arxiv.org/abs/2007.10353} {arXiv:2007.10353 [hep-th]} \BibitemShut {NoStop}%
\bibitem [{\citenamefont {Bonanno}\ \emph {et~al.}(2022)\citenamefont {Bonanno}, \citenamefont {Denz}, \citenamefont {Pawlowski},\ and\ \citenamefont {Reichert}}]{Bonanno:2021squ}%
  \BibitemOpen
  \bibfield  {author} {\bibinfo {author} {\bibfnamefont {A.}~\bibnamefont {Bonanno}}, \bibinfo {author} {\bibfnamefont {T.}~\bibnamefont {Denz}}, \bibinfo {author} {\bibfnamefont {J.~M.}\ \bibnamefont {Pawlowski}}, \ and\ \bibinfo {author} {\bibfnamefont {M.}~\bibnamefont {Reichert}},\ }\href {\doibase 10.21468/SciPostPhys.12.1.001} {\bibfield  {journal} {\bibinfo  {journal} {SciPost Phys.}\ }\textbf {\bibinfo {volume} {12}},\ \bibinfo {pages} {001} (\bibinfo {year} {2022})},\ \Eprint {http://arxiv.org/abs/2102.02217} {arXiv:2102.02217 [hep-th]} \BibitemShut {NoStop}%
\bibitem [{\citenamefont {Koch}\ and\ \citenamefont {Saueressig}()}]{Koch:2013owa}%
  \BibitemOpen
  \bibfield  {author} {\bibinfo {author} {\bibfnamefont {B.}~\bibnamefont {Koch}}\ and\ \bibinfo {author} {\bibfnamefont {F.}~\bibnamefont {Saueressig}},\ }\href {\doibase 10.1088/0264-9381/31/1/015006} {\bibfield  {journal} {\bibinfo  {journal} {\href{https://iopscience.iop.org/article/10.1088/0264-9381/31/1/015006}{"Class. Quant. Grav." \textbf{31} (2014), 015006}}\ }10.1088/0264-9381/31/1/015006},\ \Eprint {http://arxiv.org/abs/1306.1546} {arXiv:1306.1546 [hep-th]} \BibitemShut {NoStop}%
\bibitem [{\citenamefont {Platania}()}]{Platania:2019kyx}%
  \BibitemOpen
  \bibfield  {author} {\bibinfo {author} {\bibfnamefont {A.}~\bibnamefont {Platania}},\ }\href {\doibase 10.1140/epjc/s10052-019-6990-2} {\bibfield  {journal} {\bibinfo  {journal} {\href{https://link.springer.com/article/10.1140/epjc/s10052-019-6990-2}{Eur. Phys. J. C \textbf{79} (2019), no. 6, 470}}\ }10.1140/epjc/s10052-019-6990-2},\ \Eprint {http://arxiv.org/abs/1903.10411} {arXiv:1903.10411 [gr-qc]} \BibitemShut {NoStop}%
\bibitem [{\citenamefont {Reuter}\ and\ \citenamefont {Weyer}(2009)}]{Reuter:2008qx}%
  \BibitemOpen
  \bibfield  {author} {\bibinfo {author} {\bibfnamefont {M.}~\bibnamefont {Reuter}}\ and\ \bibinfo {author} {\bibfnamefont {H.}~\bibnamefont {Weyer}},\ }\href {\doibase 10.1103/PhysRevD.80.025001} {\bibfield  {journal} {\bibinfo  {journal} {Phys. Rev. D}\ }\textbf {\bibinfo {volume} {80}},\ \bibinfo {pages} {025001} (\bibinfo {year} {2009})},\ \Eprint {http://arxiv.org/abs/0804.1475} {arXiv:0804.1475 [hep-th]} \BibitemShut {NoStop}%
\bibitem [{\citenamefont {Daum}\ and\ \citenamefont {Reuter}(2009)}]{Daum:2008gr}%
  \BibitemOpen
  \bibfield  {author} {\bibinfo {author} {\bibfnamefont {J.-E.}\ \bibnamefont {Daum}}\ and\ \bibinfo {author} {\bibfnamefont {M.}~\bibnamefont {Reuter}},\ }\href {\doibase 10.1166/asl.2009.1033} {\bibfield  {journal} {\bibinfo  {journal} {Adv. Sci. Lett.}\ }\textbf {\bibinfo {volume} {2}},\ \bibinfo {pages} {255} (\bibinfo {year} {2009})},\ \Eprint {http://arxiv.org/abs/0806.3907} {arXiv:0806.3907 [hep-th]} \BibitemShut {NoStop}%
\bibitem [{\citenamefont {Bonanno}\ and\ \citenamefont {Guarnieri}(2012)}]{Bonanno:2012dg}%
  \BibitemOpen
  \bibfield  {author} {\bibinfo {author} {\bibfnamefont {A.}~\bibnamefont {Bonanno}}\ and\ \bibinfo {author} {\bibfnamefont {F.}~\bibnamefont {Guarnieri}},\ }\href {\doibase 10.1103/PhysRevD.86.105027} {\bibfield  {journal} {\bibinfo  {journal} {Phys. Rev. D}\ }\textbf {\bibinfo {volume} {86}},\ \bibinfo {pages} {105027} (\bibinfo {year} {2012})},\ \Eprint {http://arxiv.org/abs/1206.6531} {arXiv:1206.6531 [hep-th]} \BibitemShut {NoStop}%
\bibitem [{\citenamefont {Bonanno}\ \emph {et~al.}(2023)\citenamefont {Bonanno}, \citenamefont {Conti},\ and\ \citenamefont {Zappal\`a}}]{Bonanno:2023fij}%
  \BibitemOpen
  \bibfield  {author} {\bibinfo {author} {\bibfnamefont {A.}~\bibnamefont {Bonanno}}, \bibinfo {author} {\bibfnamefont {M.}~\bibnamefont {Conti}}, \ and\ \bibinfo {author} {\bibfnamefont {D.}~\bibnamefont {Zappal\`a}},\ }\href {\doibase 10.1016/j.physletb.2023.138311} {\bibfield  {journal} {\bibinfo  {journal} {Phys. Lett. B}\ }\textbf {\bibinfo {volume} {847}},\ \bibinfo {pages} {138311} (\bibinfo {year} {2023})},\ \Eprint {http://arxiv.org/abs/2309.15514} {arXiv:2309.15514 [hep-th]} \BibitemShut {NoStop}%
\bibitem [{\citenamefont {Bonanno}\ \emph {et~al.}(2020)\citenamefont {Bonanno}, \citenamefont {Eichhorn}, \citenamefont {Gies}, \citenamefont {Pawlowski}, \citenamefont {Percacci}, \citenamefont {Reuter}, \citenamefont {Saueressig},\ and\ \citenamefont {Vacca}}]{Bonanno:2020bil}%
  \BibitemOpen
  \bibfield  {author} {\bibinfo {author} {\bibfnamefont {A.}~\bibnamefont {Bonanno}}, \bibinfo {author} {\bibfnamefont {A.}~\bibnamefont {Eichhorn}}, \bibinfo {author} {\bibfnamefont {H.}~\bibnamefont {Gies}}, \bibinfo {author} {\bibfnamefont {J.~M.}\ \bibnamefont {Pawlowski}}, \bibinfo {author} {\bibfnamefont {R.}~\bibnamefont {Percacci}}, \bibinfo {author} {\bibfnamefont {M.}~\bibnamefont {Reuter}}, \bibinfo {author} {\bibfnamefont {F.}~\bibnamefont {Saueressig}}, \ and\ \bibinfo {author} {\bibfnamefont {G.~P.}\ \bibnamefont {Vacca}},\ }\href {\doibase 10.3389/fphy.2020.00269} {\bibfield  {journal} {\bibinfo  {journal} {Front. in Phys.}\ }\textbf {\bibinfo {volume} {8}},\ \bibinfo {pages} {269} (\bibinfo {year} {2020})},\ \Eprint {http://arxiv.org/abs/2004.06810} {arXiv:2004.06810 [gr-qc]} \BibitemShut {NoStop}%
\bibitem [{\citenamefont {{Mannheim}}\ and\ \citenamefont {{Kazanas}}(1989)}]{Mannheim}%
  \BibitemOpen
  \bibfield  {author} {\bibinfo {author} {\bibfnamefont {P.~D.}\ \bibnamefont {{Mannheim}}}\ and\ \bibinfo {author} {\bibfnamefont {D.}~\bibnamefont {{Kazanas}}},\ }\href {\doibase 10.1086/167623} {\bibfield  {journal} {\bibinfo  {journal} {\apj}\ }\textbf {\bibinfo {volume} {342}},\ \bibinfo {pages} {635} (\bibinfo {year} {1989})}\BibitemShut {NoStop}%
\bibitem [{\citenamefont {Cadoni}\ \emph {et~al.}(2023{\natexlab{b}})\citenamefont {Cadoni}, \citenamefont {Oi},\ and\ \citenamefont {Sanna}}]{Cadoni:2023tse}%
  \BibitemOpen
  \bibfield  {author} {\bibinfo {author} {\bibfnamefont {M.}~\bibnamefont {Cadoni}}, \bibinfo {author} {\bibfnamefont {M.}~\bibnamefont {Oi}}, \ and\ \bibinfo {author} {\bibfnamefont {A.~P.}\ \bibnamefont {Sanna}},\ }\href {\doibase 10.1007/JHEP06(2023)211} {\bibfield  {journal} {\bibinfo  {journal} {JHEP}\ }\textbf {\bibinfo {volume} {06}},\ \bibinfo {pages} {211} (\bibinfo {year} {2023}{\natexlab{b}})},\ \Eprint {http://arxiv.org/abs/2303.05557} {arXiv:2303.05557 [hep-th]} \BibitemShut {NoStop}%
\bibitem [{\citenamefont {Maldacena}\ \emph {et~al.}(1999)\citenamefont {Maldacena}, \citenamefont {Michelson},\ and\ \citenamefont {Strominger}}]{Maldacena:1998uz}%
  \BibitemOpen
  \bibfield  {author} {\bibinfo {author} {\bibfnamefont {J.~M.}\ \bibnamefont {Maldacena}}, \bibinfo {author} {\bibfnamefont {J.}~\bibnamefont {Michelson}}, \ and\ \bibinfo {author} {\bibfnamefont {A.}~\bibnamefont {Strominger}},\ }\href {\doibase 10.1088/1126-6708/1999/02/011} {\bibfield  {journal} {\bibinfo  {journal} {JHEP}\ }\textbf {\bibinfo {volume} {02}},\ \bibinfo {pages} {011} (\bibinfo {year} {1999})},\ \Eprint {http://arxiv.org/abs/hep-th/9812073} {arXiv:hep-th/9812073} \BibitemShut {NoStop}%
\bibitem [{\citenamefont {Almheiri}\ \emph {et~al.}(2020)\citenamefont {Almheiri}, \citenamefont {Hartman}, \citenamefont {Maldacena}, \citenamefont {Shaghoulian},\ and\ \citenamefont {Tajdini}}]{Almheiri:2019qdq}%
  \BibitemOpen
  \bibfield  {author} {\bibinfo {author} {\bibfnamefont {A.}~\bibnamefont {Almheiri}}, \bibinfo {author} {\bibfnamefont {T.}~\bibnamefont {Hartman}}, \bibinfo {author} {\bibfnamefont {J.}~\bibnamefont {Maldacena}}, \bibinfo {author} {\bibfnamefont {E.}~\bibnamefont {Shaghoulian}}, \ and\ \bibinfo {author} {\bibfnamefont {A.}~\bibnamefont {Tajdini}},\ }\href {\doibase 10.1007/JHEP05(2020)013} {\bibfield  {journal} {\bibinfo  {journal} {JHEP}\ }\textbf {\bibinfo {volume} {05}},\ \bibinfo {pages} {013} (\bibinfo {year} {2020})},\ \Eprint {http://arxiv.org/abs/1911.12333} {arXiv:1911.12333 [hep-th]} \BibitemShut {NoStop}%
\bibitem [{\citenamefont {Penington}\ \emph {et~al.}(2022)\citenamefont {Penington}, \citenamefont {Shenker}, \citenamefont {Stanford},\ and\ \citenamefont {Yang}}]{Penington:2019kki}%
  \BibitemOpen
  \bibfield  {author} {\bibinfo {author} {\bibfnamefont {G.}~\bibnamefont {Penington}}, \bibinfo {author} {\bibfnamefont {S.~H.}\ \bibnamefont {Shenker}}, \bibinfo {author} {\bibfnamefont {D.}~\bibnamefont {Stanford}}, \ and\ \bibinfo {author} {\bibfnamefont {Z.}~\bibnamefont {Yang}},\ }\href {\doibase 10.1007/JHEP03(2022)205} {\bibfield  {journal} {\bibinfo  {journal} {JHEP}\ }\textbf {\bibinfo {volume} {03}},\ \bibinfo {pages} {205} (\bibinfo {year} {2022})},\ \Eprint {http://arxiv.org/abs/1911.11977} {arXiv:1911.11977 [hep-th]} \BibitemShut {NoStop}%
\bibitem [{\citenamefont {Bonanno}\ and\ \citenamefont {Reuter}(2007)}]{Bonanno:2007wg}%
  \BibitemOpen
  \bibfield  {author} {\bibinfo {author} {\bibfnamefont {A.}~\bibnamefont {Bonanno}}\ and\ \bibinfo {author} {\bibfnamefont {M.}~\bibnamefont {Reuter}},\ }\href {\doibase 10.1088/1475-7516/2007/08/024} {\bibfield  {journal} {\bibinfo  {journal} {JCAP}\ }\textbf {\bibinfo {volume} {08}},\ \bibinfo {pages} {024} (\bibinfo {year} {2007})},\ \Eprint {http://arxiv.org/abs/0706.0174} {arXiv:0706.0174 [hep-th]} \BibitemShut {NoStop}%
\bibitem [{\citenamefont {Cadoni}\ and\ \citenamefont {Sanna}(2023)}]{Cadoni:2023nrm}%
  \BibitemOpen
  \bibfield  {author} {\bibinfo {author} {\bibfnamefont {M.}~\bibnamefont {Cadoni}}\ and\ \bibinfo {author} {\bibfnamefont {A.~P.}\ \bibnamefont {Sanna}},\ }\href {\doibase 10.1088/1361-6382/acde3c} {\bibfield  {journal} {\bibinfo  {journal} {Class. Quant. Grav.}\ }\textbf {\bibinfo {volume} {40}},\ \bibinfo {pages} {145012} (\bibinfo {year} {2023})},\ \Eprint {http://arxiv.org/abs/2302.06401} {arXiv:2302.06401 [gr-qc]} \BibitemShut {NoStop}%
\bibitem [{\citenamefont {Cadoni}\ \emph {et~al.}(2022)\citenamefont {Cadoni}, \citenamefont {Oi},\ and\ \citenamefont {Sanna}}]{Cadoni:2022chn}%
  \BibitemOpen
  \bibfield  {author} {\bibinfo {author} {\bibfnamefont {M.}~\bibnamefont {Cadoni}}, \bibinfo {author} {\bibfnamefont {M.}~\bibnamefont {Oi}}, \ and\ \bibinfo {author} {\bibfnamefont {A.~P.}\ \bibnamefont {Sanna}},\ }\href {\doibase 10.1103/PhysRevD.106.024030} {\bibfield  {journal} {\bibinfo  {journal} {Phys. Rev. D}\ }\textbf {\bibinfo {volume} {106}},\ \bibinfo {pages} {024030} (\bibinfo {year} {2022})},\ \Eprint {http://arxiv.org/abs/2204.09444} {arXiv:2204.09444 [gr-qc]} \BibitemShut {NoStop}%
\bibitem [{\citenamefont {Belgacem}\ \emph {et~al.}(2018)\citenamefont {Belgacem}, \citenamefont {Dirian}, \citenamefont {Foffa},\ and\ \citenamefont {Maggiore}}]{Belgacem:2017cqo}%
  \BibitemOpen
  \bibfield  {author} {\bibinfo {author} {\bibfnamefont {E.}~\bibnamefont {Belgacem}}, \bibinfo {author} {\bibfnamefont {Y.}~\bibnamefont {Dirian}}, \bibinfo {author} {\bibfnamefont {S.}~\bibnamefont {Foffa}}, \ and\ \bibinfo {author} {\bibfnamefont {M.}~\bibnamefont {Maggiore}},\ }\href {\doibase 10.1088/1475-7516/2018/03/002} {\bibfield  {journal} {\bibinfo  {journal} {JCAP}\ }\textbf {\bibinfo {volume} {03}},\ \bibinfo {pages} {002} (\bibinfo {year} {2018})},\ \Eprint {http://arxiv.org/abs/1712.07066} {arXiv:1712.07066 [hep-th]} \BibitemShut {NoStop}%
\bibitem [{\citenamefont {Del~Piano}\ \emph {et~al.}(2024{\natexlab{a}})\citenamefont {Del~Piano}, \citenamefont {Hohenegger},\ and\ \citenamefont {Sannino}}]{Sannino:2023fiw}%
  \BibitemOpen
  \bibfield  {author} {\bibinfo {author} {\bibfnamefont {M.}~\bibnamefont {Del~Piano}}, \bibinfo {author} {\bibfnamefont {S.}~\bibnamefont {Hohenegger}}, \ and\ \bibinfo {author} {\bibfnamefont {F.}~\bibnamefont {Sannino}},\ }\href {\doibase 10.1103/PhysRevD.109.024045} {\bibfield  {journal} {\bibinfo  {journal} {Phys. Rev. D}\ }\textbf {\bibinfo {volume} {109}},\ \bibinfo {pages} {024045} (\bibinfo {year} {2024}{\natexlab{a}})},\ \Eprint {http://arxiv.org/abs/2307.13489} {arXiv:2307.13489 [gr-qc]} \BibitemShut {NoStop}%
\bibitem [{\citenamefont {Del~Piano}\ \emph {et~al.}(2024{\natexlab{b}})\citenamefont {Del~Piano}, \citenamefont {Hohenegger},\ and\ \citenamefont {Sannino}}]{Sannino:2024gvw}%
  \BibitemOpen
  \bibfield  {author} {\bibinfo {author} {\bibfnamefont {M.}~\bibnamefont {Del~Piano}}, \bibinfo {author} {\bibfnamefont {S.}~\bibnamefont {Hohenegger}}, \ and\ \bibinfo {author} {\bibfnamefont {F.}~\bibnamefont {Sannino}},\ }\href {\doibase 10.1140/epjc/s10052-024-13609-5} {\bibfield  {journal} {\bibinfo  {journal} {Eur. Phys. J. C}\ }\textbf {\bibinfo {volume} {84}},\ \bibinfo {pages} {1273} (\bibinfo {year} {2024}{\natexlab{b}})},\ \Eprint {http://arxiv.org/abs/2403.12679} {arXiv:2403.12679 [gr-qc]} \BibitemShut {NoStop}%
\bibitem [{\citenamefont {Bonanno}\ \emph {et~al.}(2018)\citenamefont {Bonanno}, \citenamefont {Platania},\ and\ \citenamefont {Saueressig}}]{Bonanno:2018gck}%
  \BibitemOpen
  \bibfield  {author} {\bibinfo {author} {\bibfnamefont {A.}~\bibnamefont {Bonanno}}, \bibinfo {author} {\bibfnamefont {A.}~\bibnamefont {Platania}}, \ and\ \bibinfo {author} {\bibfnamefont {F.}~\bibnamefont {Saueressig}},\ }\href {\doibase 10.1016/j.physletb.2018.06.047} {\bibfield  {journal} {\bibinfo  {journal} {Phys. Lett. B}\ }\textbf {\bibinfo {volume} {784}},\ \bibinfo {pages} {229} (\bibinfo {year} {2018})},\ \Eprint {http://arxiv.org/abs/1803.02355} {arXiv:1803.02355 [gr-qc]} \BibitemShut {NoStop}%
\bibitem [{\citenamefont {Biemans}\ \emph {et~al.}(2017)\citenamefont {Biemans}, \citenamefont {Platania},\ and\ \citenamefont {Saueressig}}]{Biemans:2017zca}%
  \BibitemOpen
  \bibfield  {author} {\bibinfo {author} {\bibfnamefont {J.}~\bibnamefont {Biemans}}, \bibinfo {author} {\bibfnamefont {A.}~\bibnamefont {Platania}}, \ and\ \bibinfo {author} {\bibfnamefont {F.}~\bibnamefont {Saueressig}},\ }\href {\doibase 10.1007/JHEP05(2017)093} {\bibfield  {journal} {\bibinfo  {journal} {JHEP}\ }\textbf {\bibinfo {volume} {05}},\ \bibinfo {pages} {093} (\bibinfo {year} {2017})},\ \Eprint {http://arxiv.org/abs/1702.06539} {arXiv:1702.06539 [hep-th]} \BibitemShut {NoStop}%
\bibitem [{\citenamefont {Alkofer}\ and\ \citenamefont {Saueressig}(2018)}]{Alkofer:2018fxj}%
  \BibitemOpen
  \bibfield  {author} {\bibinfo {author} {\bibfnamefont {N.}~\bibnamefont {Alkofer}}\ and\ \bibinfo {author} {\bibfnamefont {F.}~\bibnamefont {Saueressig}},\ }\href {\doibase 10.1016/j.aop.2018.07.017} {\bibfield  {journal} {\bibinfo  {journal} {Annals Phys.}\ }\textbf {\bibinfo {volume} {396}},\ \bibinfo {pages} {173} (\bibinfo {year} {2018})},\ \Eprint {http://arxiv.org/abs/1802.00498} {arXiv:1802.00498 [hep-th]} \BibitemShut {NoStop}%
\bibitem [{\citenamefont {Percacci}\ and\ \citenamefont {Perini}(2004)}]{Percacci:2004sb}%
  \BibitemOpen
  \bibfield  {author} {\bibinfo {author} {\bibfnamefont {R.}~\bibnamefont {Percacci}}\ and\ \bibinfo {author} {\bibfnamefont {D.}~\bibnamefont {Perini}},\ }\href {\doibase 10.1088/0264-9381/21/22/002} {\bibfield  {journal} {\bibinfo  {journal} {Class. Quant. Grav.}\ }\textbf {\bibinfo {volume} {21}},\ \bibinfo {pages} {5035} (\bibinfo {year} {2004})},\ \Eprint {http://arxiv.org/abs/hep-th/0401071} {arXiv:hep-th/0401071} \BibitemShut {NoStop}%
\bibitem [{\citenamefont {Cadoni}\ \emph {et~al.}(2023{\natexlab{c}})\citenamefont {Cadoni}, \citenamefont {De~Laurentis}, \citenamefont {De~Martino}, \citenamefont {Della~Monica}, \citenamefont {Oi},\ and\ \citenamefont {Sanna}}]{Cadoni:2022vsn}%
  \BibitemOpen
  \bibfield  {author} {\bibinfo {author} {\bibfnamefont {M.}~\bibnamefont {Cadoni}}, \bibinfo {author} {\bibfnamefont {M.}~\bibnamefont {De~Laurentis}}, \bibinfo {author} {\bibfnamefont {I.}~\bibnamefont {De~Martino}}, \bibinfo {author} {\bibfnamefont {R.}~\bibnamefont {Della~Monica}}, \bibinfo {author} {\bibfnamefont {M.}~\bibnamefont {Oi}}, \ and\ \bibinfo {author} {\bibfnamefont {A.~P.}\ \bibnamefont {Sanna}},\ }\href {\doibase 10.1103/PhysRevD.107.044038} {\bibfield  {journal} {\bibinfo  {journal} {Phys. Rev. D}\ }\textbf {\bibinfo {volume} {107}},\ \bibinfo {pages} {044038} (\bibinfo {year} {2023}{\natexlab{c}})},\ \Eprint {http://arxiv.org/abs/2211.11585} {arXiv:2211.11585 [gr-qc]} \BibitemShut {NoStop}%
\end{thebibliography}%

\end{document}